\documentclass{aa}  

\usepackage{graphicx}
\usepackage{caption}
\usepackage{subcaption}
\usepackage{verbatim}
\usepackage{lipsum}
\usepackage{color}

\usepackage[switch, displaymath, mathlines]{lineno}

\usepackage[varg]{txfonts}
\begin{document} 

\title{Design and performance of dual-polarization lumped-element kinetic inductance detectors for millimeter-wave polarimetry}

   \subtitle{ }

   \author{H.~McCarrick\inst{\ref{columbia}}
          \and
          G.~Jones\inst{\ref{columbia}}
          \and  
          B.~R.~Johnson\inst{\ref{columbia}}
          \and 
          M.~H.~Abitbol\inst{\ref{columbia}}
          \and
          P.~A.~R.~Ade\inst{\ref{cardiff}}
          \and 
          S.~Bryan\inst{\ref{asu}}
          \and  
          P.~Day\inst{\ref{jpl}}
          \and 
          T.~Essinger-Hileman\inst{\ref{goddard}}
          \and
          D.~Flanigan\inst{\ref{columbia}}
          \and 
          H.~G.~Leduc\inst{\ref{jpl}}
          \and 
          M.~Limon\inst{\ref{columbia}}
          \and 
          P.~Mauskopf\inst{\ref{asu}}
          \and 
          A.~Miller\inst{\ref{usc}}
          \and
          C.~Tucker\inst{\ref{cardiff}} 
          }
    
   \institute{Department of Physics, 
   			  Columbia University,
              New York, NY 10027, USA\\
              \email{hlm2124@columbia.edu}\label{columbia}
         \and 
        	 School of Physics and Astronomy, 
             Cardiff University, 
             Cardiff, Wales CF24 3AA, UK \label{cardiff}
         \and
             School of Earth and Space Exploration, 
             Arizona State University, 
             Tempe, AZ 85287, USA \label{asu}
         \and 
          	 Jet Propulsion Laboratory, 
             Pasadena, CA 91109, USA \label{jpl}
         \and 
         	 Goddard Space Flight Center, 
             Greenbelt, MD 20771, USA \label{goddard}
         \and
        	Department of Physics and Astronomy, 
         	University of Southern California, 
         	Los Angeles, CA 90089, USA \label{usc}
             }
             
   \date{Accepted 15 November 2017}

 
  \abstract
   {}
   {Lumped-element kinetic inductance detectors (LEKIDs) are an attractive technology for millimeter-wave observations that require large arrays of extremely low-noise detectors.
   We designed, fabricated and characterized 64-element (128 LEKID) arrays of horn-coupled, dual-polarization LEKIDs optimized for ground-based CMB polarimetry.
   Our devices are sensitive to two orthogonal polarizations in a single spectral band centered on 150~GHz with $\Delta \nu/ \nu = 0.2$. 
   The $65 \times 65$~mm square arrays are designed to be tiled into the focal plane of an optical system.
   We demonstrate the viability of these dual-polarization LEKIDs with laboratory measurements.}
   {The LEKID modules are tested with an FPGA-based readout system in a sub-kelvin cryostat that uses a two-stage adiabatic demagnetization refrigerator. 
   The devices are characterized using a blackbody and a millimeter-wave source. 
   The polarization properties are measured with a cryogenic stepped half-wave plate. 
   We measure the resonator parameters and the detector sensitivity, noise spectrum, dynamic range, and polarization response.}
   {The resonators have internal quality factors approaching $1 \times 10^{6}$. 
   The detectors have uniform response between orthogonal polarizations and a large dynamic range. 
   The detectors are photon-noise limited above 1~pW of absorbed power. 
   The noise-equivalent temperatures under a 3.4~K blackbody load are $<100~\mu\mathrm{K}\sqrt{\mathrm{s}}$. 
   The polarization fractions of detectors sensitive to orthogonal polarizations are $>80 \%$. 
   The entire array is multiplexed on a single readout line, demonstrating a multiplexing factor of 128. 
   The array and readout meet the requirements for 4 arrays to be read out simultaneously for a multiplexing factor of 512.}
   {This laboratory study demonstrates the first dual-polarization LEKID array optimized specifically for CMB polarimetry and shows the readiness of the detectors for on-sky observations.}
               
   \titlerunning{Dual-polarization LEKIDs for millimeter-wave polarimetry}
   \maketitle


\section{Introduction}
\label{sec:introduction}

   \begin{figure*}[t!]
   \centering
         \includegraphics[width=\textwidth]{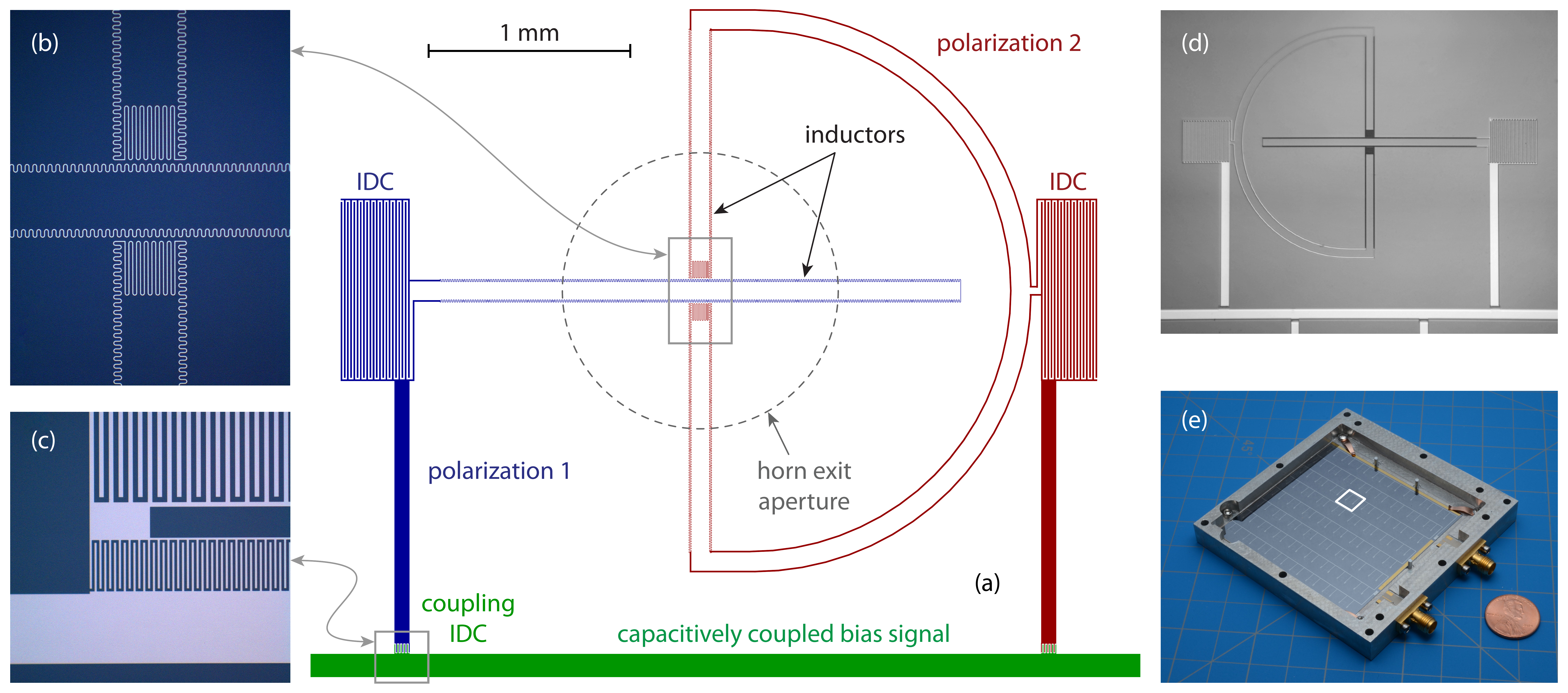}
         \caption{a) Schematic of a dual-polarization LEKID array element. Each array element consists of two LEKID resonators sensitive to orthogonal polarizations. The resonators sensitive to the two polarizations are shown in red and in blue. Each resonator consists of an inductor and an interdigitated capacitor (IDC). The LEKID absorbs the incident radiation in the inductor. The detectors are horn coupled and the aperture of the waveguide at the end of the horn is shown as the dashed gray line. 
         b) Photograph of the millimeter-wave absorbers for a single array element. The inductors are `wiggled' to increase the active volume of the absorber and thus the dynamic range of the detectors. The meanders at the end of polarization 2 (shown in red in (a)) are used to decrease the absorption of cross-polarization. 
         c) Photograph of the coupler and resonator capacitor. This photograph is of a $\sim$100~MHz resonator that has a large capacitor that is visible in the top of the photo. All detectors are capacitively coupled to, and read out on, a single transmission line. 
         d) Photograph of a single dual-polarization LEKID array element with $\sim$~1~GHz resonance frequencies.
         e) Photograph of a dual-polarization LEKID array. The white box highlights a single array element, as shown in (d). A single array consists of 64 elements (128 LEKIDs) fabricated on a 100~mm diameter silicon wafer. The arrays are diced into squares and designed to be tiled to fill the focal plane of a telescope.
         }
         \label{fig:kid_design}
    \end{figure*}

Lumped-element kinetic inductance detectors (LEKIDs) are planar, superconducting LC resonators that are also photon absorbers. An absorbed photon with an energy greater than $2 \Delta$, where $\Delta$ is the superconducting gap, will break Cooper pair(s) increasing the quasiparticle density $n_{\mathrm{qp}}$, dissipation, and kinetic inductance $L_\mathrm{k}$~\citep{day_2003,doyle_2008}. This results in a shift in the resonator quality factor $Q$ and resonance frequency $f_\mathrm{0}$, which is read out by monitoring a probe tone that is driving the device at its nominal resonance frequency. Each LEKID has a unique resonance frequency set by the geometry of the capacitor. This architecture naturally allows for frequency multiplexing because an array of LEKIDs can be read out on a single transmission line. 

Kinetic inductance detectors (KIDs) have been developed for a wide range of wavelengths~\citep{calvo_2016, mccarrick_2014, Mazin2013, swenson_2012}. Significant advances have been made in developing dual-polarization and multi-chroic KIDS~\citep{Dober_2016, mccarrick_2016b, johnson_2016}, demonstrating photon-noise dominated sensitivity~\citep{bueno2016, flanigan_2016a, Hubmayr2014, mauskopf14, janssen2013}, determining space-readiness in the far infrared (FIR)~\citep{baselmans2016}, and deploying arrays for astrophysical observations~\citep{adam_2017, nikatsz13, szypryt2014}. 

In this paper we describe the design and measured performance of dual-polarization LEKIDs optimized for cosmic microwave background (CMB) studies.
CMB experiments are being designed to search for the divergence-free component of the polarization signal, often referred to as the B-mode signal, which would provide strong evidence for inflation after the Big Bang~\citep{knox_2002, seljak_1997, kamionkowski_1997}.
The anticipated primordial B-mode signal is faint when compared with the unavoidable photon noise in the CMB itself, and current instruments already use photon-noise limited detectors~\citep{simons_array_2016,advanced_actpol_2016,spt-3g_2016,class_2016,bicep3_array_2016}.
Therefore, to increase instrument sensitivity going forward, the number of detectors must be increased, so scalable detector technologies with high multiplexing factors are needed for next-generation CMB experiments~\citep{cmb_s4_technology}.
LEKIDs could be an attractive option for these next-generation experiments, so we conducted the laboratory study described in this paper to investigate their suitability.
In particular, we focused on measuring the sensitivity and polarization selectivity of our design to see if it is viable, and we tested a multiplexing factor of 128 that advances the state-of-the-art for CMB experiments.

The paper is organized as follows: In Section 2.1, we present the detector requirements and subsequent design. In Section 2.2, we discuss the optical coupling and array design. In Section 2.3, we describe the experimental system. In Section 3, we present the tests undertaken to characterize the detectors including measurements of the resonator parameters, optical response, dynamic range, noise, and polarization selectivity. In Section 4, we discuss the measurement results. 


   \begin{figure*}[t!]
   \centering
         \includegraphics[width=\textwidth]{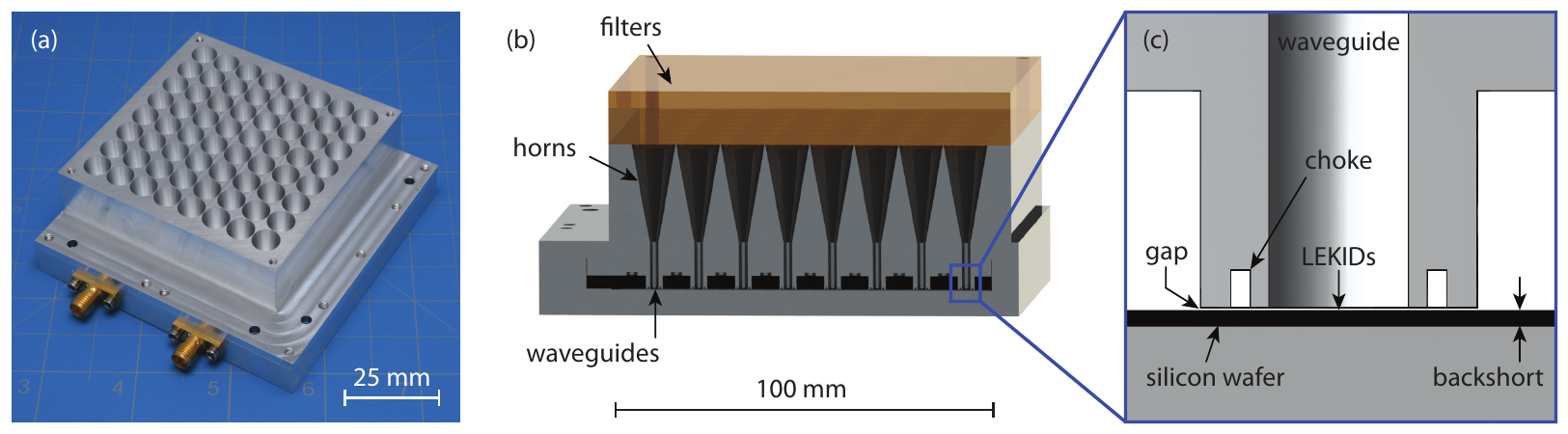}
         \caption{
         a) Photograph of the array module with the horn apertures visible. Conical horns are used to couple the incoming radiation to the detectors.  
         b) Cross-sectional view of the horn array. Each horn feeds a cylindrical waveguide. At the waveguide exit a choke is used to optimize the coupling between the horn and the detector, as shown in (c). The LEKIDs are front-side illuminated. The detectors are fabricated on a 160~$\mu$m thick silicon wafer. The wafer thickness sets the $\lambda$/4 backshort distance and the back side of the wafer is covered with 100~nm thick Al, which acts as a backshort and ground plane. There are pockets between the waveguides into which the spring-loaded pins that press down on the detector array are epoxied.
         c) Cross-sectional view of the coupling between the horn and one LEKID array element.}
         \label{fig:design}
    \end{figure*}


\section{Methods}
\subsection{Detector requirements and design}
\label{sec:detector_design}

The dual-polarization LEKIDs are laid out in an $8 \times 8$ square array with a 7.8~mm pitch (see Fig.~\ref{fig:kid_design}e). 
Each array element consists of two resonators for a total of 128 LEKIDs per array. 
The array element design is shown in Fig.~\ref{fig:kid_design}. 
The two resonators are sensitive to orthogonal polarizations for observation in a single spectral band centered on 150~GHz. 
Each resonator consists of an inductor connected in parallel with an interdigitated capacitor (IDC).
The inductors for each pixel are identical across the array.
Each IDC has a unique capacitance and thus each resonator has a unique $f_\mathrm{0}$.
Therefore the detectors naturally lend themselves to frequency multiplexing.

%
The arrays are fabricated on high-resistivity (> 10~k$\Omega$cm) silicon wafers.
After the detector processing steps are completed, the silicon is thinned to 160~$\mu$m by grinding the back side of the wafer. 
As the final step, the arrays are diced into squares, which are designed to be tiled. 
The total number of detectors in each array is therefore determined by the choice of pitch and the decision to use 100~mm diameter wafers as the substrate. 
%

The resonators are capacitively coupled to a 50~$\Omega$ microstrip transmission line. 
We chose to metalize the back side of the wafer with a 100~nm thick aluminum (Al) film to provide a ground plane for both the microstrip and the resonators and to optimize the millimeter-wave coupling (see Fig.~\ref{fig:design} and Sec.~\ref{sec:coupling}). 
The detectors and transmission line are made out of a single Al film that is 25~nm thick. 
Wirebonds connect the Al microstrip to the rest of the readout chain (see Sec.~\ref{sec:readout}). 
We measured the critical temperature of the film to be 1.4~K. 
%

%
The absorbing element of the detector is the inductor. 
In our design, the geometry of one inductor is a single, long trace in the shape of a hairpin, while the orthogonal inductor is composed of two hairpins.
We chose this architecture so the array can be fabricated with a single film without crossovers. 

The detectors are designed for ground-based CMB experiments which have an expected loading of 1 to 10~pW. 
We started with straight inductors~\citep{bryan_2015, mccarrick_2016b} but found for this absorber geometry the resonator quality factor degraded too rapidly as a function of absorbed power $P$.
To retain sufficiently high $Q$ under ground based loading conditions, we increased the volume $V_\mathrm{L}$ of the inductor. 
{Solely increasing the film thickness would decrease the surface impedance of the absorber, causing poor coupling to the incoming radiation.}
{Therefore,} $V_\mathrm{L}$ was tuned by `wiggling' the trace in alternating semi-circles as shown in the photograph in Fig.~\ref{fig:kid_design}b.
Wiggling the inductors allows us to add more volume by increasing both the length and film thickness while maintaining a similar effective surface impedance to the millimeter-wave radiation. 
The wiggles increase the length of the inductors as compared with a straight meander by a factor of $\mathrm{\pi}$/2.  
To maintain the surface impedance needed for high optical coupling, we made the film thickness 25~nm. 
The detector response $\mathrm{d}x/\mathrm{d}P$ is defined as the ratio of the change in the fractional frequency shift 
\begin{align}
\label{eq:x}
x = (f - f_{\mathrm{0}})/f_{\mathrm{0}}
\end{align}
to the change in absorbed power.
As a result of the increased volume, we expect $\mathrm{d}x/\mathrm{d}P$ to decrease as compared with the straight inductor for a given $P$. 

\begin{figure}[t!]
\centering
\includegraphics[width=\linewidth]{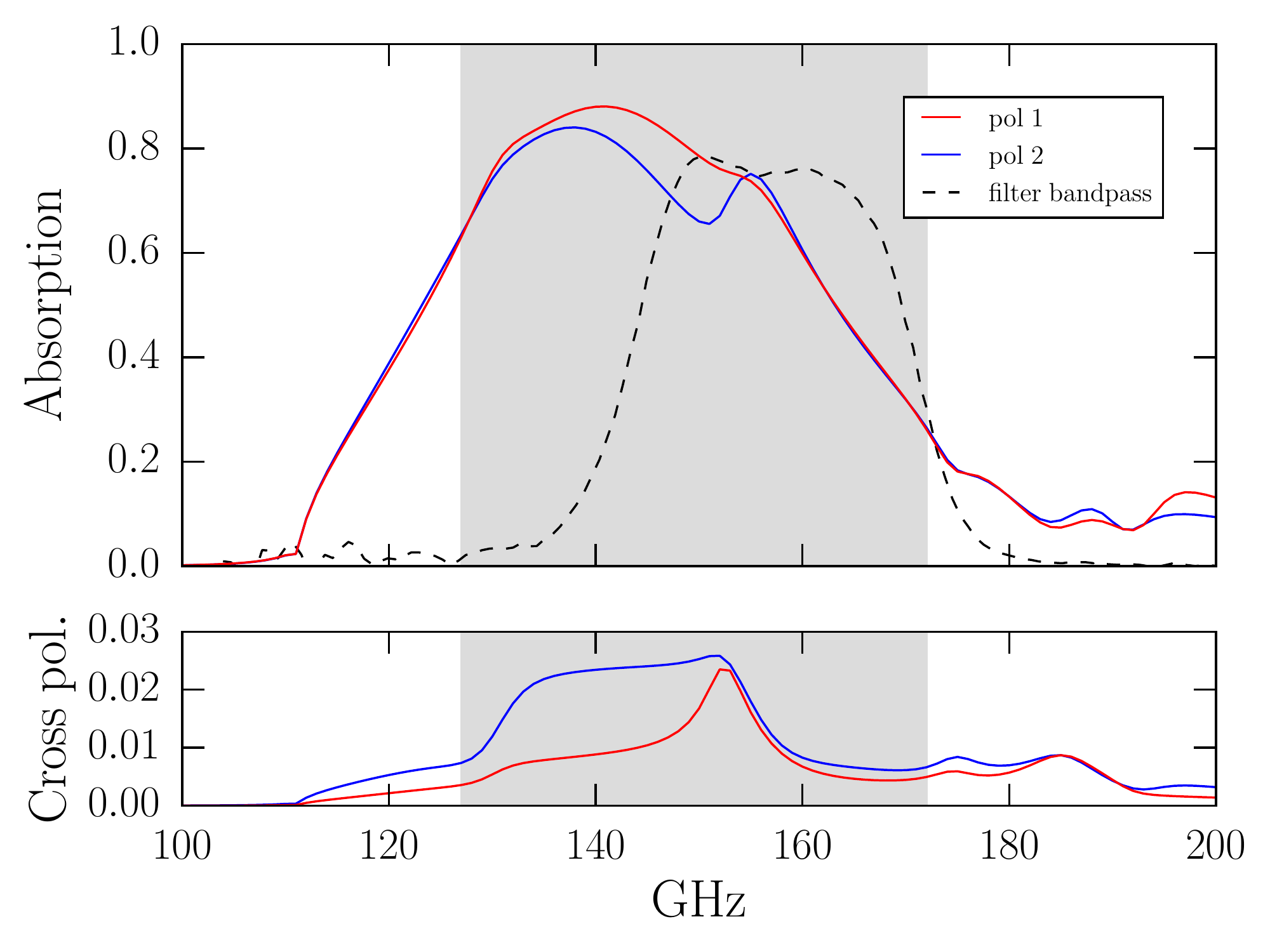}
\caption{Plot of the simulated absorption spectra of the detectors sensitive to orthogonal polarizations. The shaded region shows the maximum spectral band the detectors could cover. The dotted black line shows the measured spectrum of the filter that defines the bandpass for the experiments described in this paper. The bottom plot shows the cross polarization absorption of the detectors. 
}
\label{fig:hfss}%
    \end{figure}

The inductors are naturally polarization sensitive, preferentially absorbing radiation with the E-field aligned to the inductor trace. 
Electromagnetic simulations\footnote{ANSYS Electronics Desktop 2016} were used to optimize the design so that the two polarizations have similar absorption spectra.
The simulations included the detector and array design starting from the waveguide (see Sec.~\ref{sec:coupling}) through the array backshort (see Fig.~\ref{fig:design}).
We found it necessary to meander the end of the polarization-1 inductor (see Fig.~\ref{fig:kid_design}) to minimize absorption of the cross-polarization. 
The simulated absorption spectra and cross-polarization spectra are shown in Fig.~\ref{fig:hfss}.
%

We have designed arrays with resonance frequencies in the 100--200~MHz or 800--1200~MHz bands. 
The arrays are fabricated with a stepper.
Each stepper field contains a $2 \times 2$  sub-array of dual-polarization LEKIDs, so an array is comprised of 16 stepper fields. 
The stepper fields are exposed in a $4 \times 4$ array across the wafer to make the full $8 \times 8$ array of dual-polarization LEKIDs.  
Each stepper field is exposed in two steps. First, the fixed portion of the design, including the inductors, coupling capacitors, transmission line segments, and half of each IDC structure, is exposed identically for each stepper field. A second stepper field including the second half of the IDC structures is then exposed with a relative offset that varies across the wafer. The capacitor lengths set by this relative offset increase linearly across the wafer, resulting in a weakly quadratic frequency spacing across the readout bandwidth. 
Within each stepper field, the resonance frequencies of the detectors are spaced maximally far apart over the array readout bandwidth to decrease the likelihood of adjacent resonators coupling to one another in frequency space~\citep{noroozian_2012}. 
The resonance frequencies for the two polarizations are separated into a high and low band within the readout bandwidth of the array. 

Our LEKID readout system provides a bandwidth of 250~MHz at baseband frequencies (0--250~MHz) or 500 MHz when using an IQ mixer to target readout frequencies in the range 0.5--4.0~GHz.
Although we are demonstrating a multiplexing factor of 128 here, our design goal is to read out 4~modules with a single readout system so we are designing for a future multiplexing factor of 512. 
To avoid collisions between detectors, we choose a resonance frequency spacing of 10 times the resonance width.
This choice means the $Q$ of the resonators needs to be $> 10^{4}$. 
The resonator quality factor $Q$ is determined as 
\begin{align}
\label{eq:qs}
Q^{-1} = Q_\mathrm{i}^{-1} + Q_\mathrm{c}^{-1},
\end{align}
where $Q_{\mathrm{i}}$ is the internal quality factor and $Q_\mathrm{c}$ is the coupling quality factor. 
We chose $Q_\mathrm{c}$ to be approximately $2 \times 10^{4}$ which is set by the value of the coupling capacitor (see Fig.~\ref{fig:kid_design}). 
These fabricated devices have $Q_\mathrm{i} \sim 10^{5}$ under an optical load, which is higher than we expected, so the coupling quality factor $Q_{\mathrm{c}}$ predominately sets the total $Q$ as $Q_{\mathrm{i}} \gg Q_{\mathrm{C}}$.
Ideally, $Q_\mathrm{i} = Q_\mathrm{c}$ under the desired optical load~\citep{zmu}, so in future iterations of the design the coupling capacitance will be decreased.


\begin{figure*}[t!]
\centering
\includegraphics[width=\textwidth]{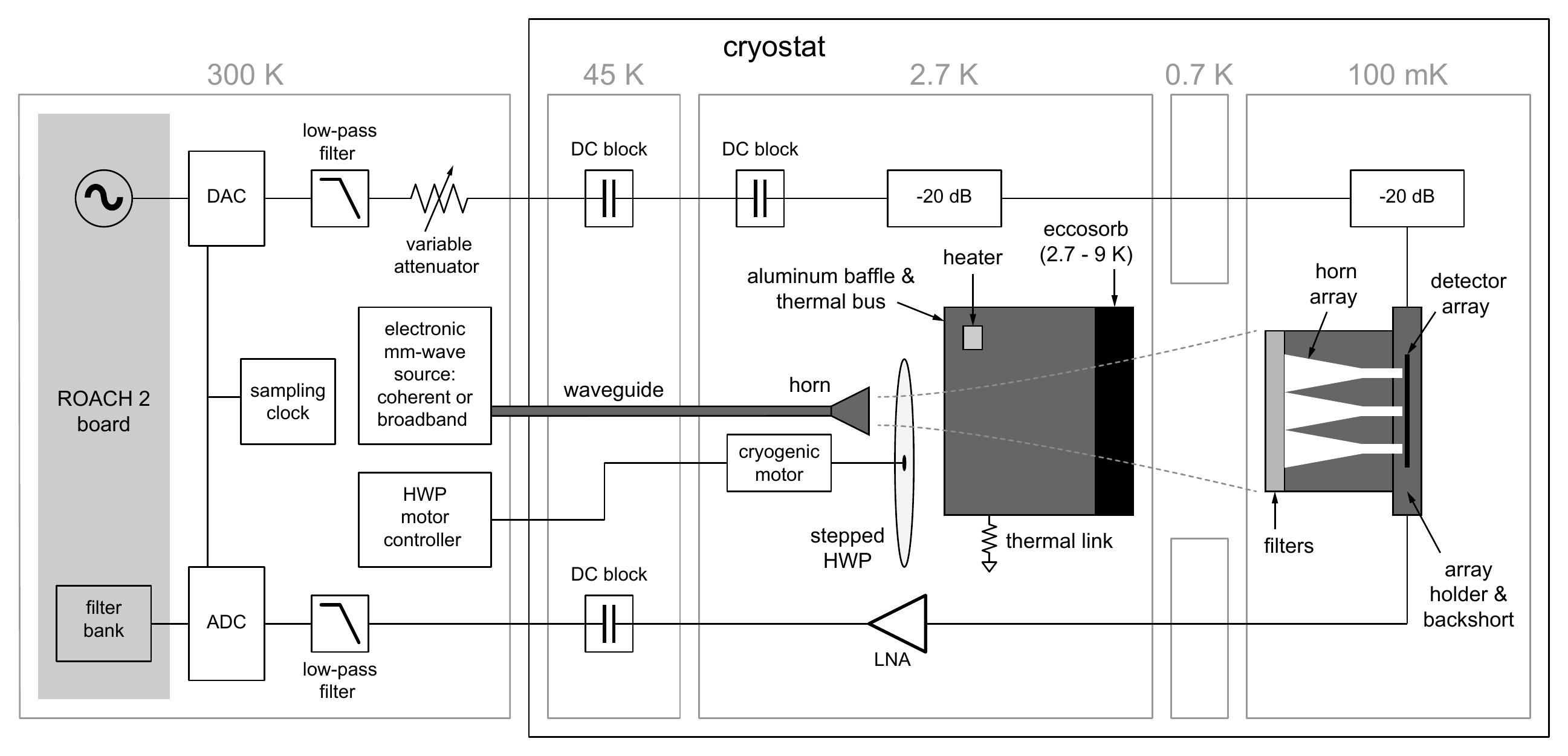}
\caption{Schematic of the LEKID test system. The LEKID readout is based on the ROACH 2 board. The 100--200~MHz resonance frequencies of the LEKIDs allow us to use the ADC/DAC at the baseband frequencies. For arrays with higher resonance frequency KIDs, we use IQ mixers which are not shown here~\citep{johnson_2016}. The entire array is read out on a single pair of coaxial lines. The millimeter-wave (MMW) sources are also shown. A half-wave plate (HWP) mounted at 2.7~K and rotated with a cryogenic stepper motor modulates the MMW radiation that enters the cryostat through a WR6 waveguide. The radiation (140--165~GHz) can either be single frequency (coherent) or broadband (incoherent)~\citep{flanigan_2016a}. Additionally, there is a blackbody source made from Eccosorb, the temperature of which can be controlled via a heater resistor.}
              \label{fig:setup}%
    \end{figure*}

\subsection{Detector array design}
\label{sec:coupling}
The absorption band is defined by metal mesh filters mounted in front of the horn apertures. 
The bandwidth is defined by a band-pass filter with a measured spectrum from 140--170 GHz as plotted in Fig.~\ref{fig:hfss}. 
For future tests or observations, it would be advantageous to use a wider bandpass filter which more closely matches the absorption spectra of the detectors.
Additional low-pass and high-pass filters suppress out-of-band spectral leaks.  

The detectors are fed with conicals horns that have a 7.8~mm aperture and a $20\degr$ flare angle (see Fig.~\ref{fig:design}). 
One horn feeds one array element (two LEKIDs) so the horns are arranged an $8 \times 8$ square array with a 7.8~mm pitch.
The horns are designed to couple to an F/2.5 optical system, which means the aperture diameter is 1.6~F$\lambda$ at 150~GHz. 
%

The horn flare feeds a cylindrical waveguide with a 1.6~mm diameter. 
This waveguide diameter has a cut-on frequency below the low-frequency edge of the spectral band defined by the band-pass filter. 
We chose this diameter because it works well with a profiled-horn design we are developing. 
Future LEKID modules will use these profiled horns because simulations show the horn beam is more circular. 
A choke at the waveguide output optimizes detector coupling and minimizes the lateral leakage of fields that can produce crosstalk between detectors.
A vacuum gap of 30~$\mu$m then precedes the LEKIDs.
The 160~$\mu$m wafer thickness sets the $\lambda/4$ backshort distance. 
The metalized ground plane on the backside of the wafer also acts as the backshort, increasing the absorption efficiency of each LEKID. 

The LEKID array is mounted to the bottom part of a two-part aluminum package, the top of which is the horn array. 
The detector array is edge aligned against dowel pins mounted in the bottom of the package. 
The same pins are used to co-align the horn array in the package lid with the LEKID array in the package bottom.
The array is held in place by three beryllium-copper clips positioned in corners of the LEKID array.
Small spring-loaded pins\footnote{Mill-Max Manufacturing Corp. ED9000-ND} epoxied\footnote{3M Scotch-Weld Epoxy Adhesive DP420 Black} into the package lid press down on the LEKID array in ten positions when the module is assembled. 
These spring-loaded pins help suppress vibrations in the array, which can produce features in the noise spectra (see Sec.~\ref{sec:noise}). 
To improve the heat sinking of the silicon substrate, gold bars were patterned on the edge of the detector array and wirebonds connect these bars to the package bottom.

\subsection{Experimental system}

\subsubsection{Cryogenics}
The experimental system is schematically depicted in Fig.~\ref{fig:setup}.
The detectors are mounted inside a cryostat\footnote{DRC-102 Cryostat System made by STAR Cryoelectronics} that uses a two-stage adiabatic demagnetization refrigerator (ADR) backed by a two-stage pulse tube cooler (PTC). 
The PTC provides 45~K and 2.7~K temperature stages, while the ADR provides a 0.7~K stage and a variable 60--300~mK stage. 
The detectors are mounted on the variable, sub-kelvin stage.  
%
\subsubsection{Readout}
\label{sec:readout}
The readout system used for this study is based on the open-source ROACH-2 board, which hosts a Xilinx Virtex-6 FPGA. 
The ROACH-2 board\footnote{Digicom Electronics} is combined with an ADC/DAC daughter board\footnote{Techne Instruments} that provides two 12-bit ADCs and two 16-bit DACs, each capable of synthesizing and analyzing signals with 250~MHz of bandwidth.

We have designed two varieties of LEKID arrays.
For one variety, the frequencies are designed to be in the range 100--200~MHz and the other in the range 800--1200~MHz.
The signals used to read out the lower-frequency 100--200~MHz resonators can be directly synthesized and analyzed using the ADC/DAC board. 
In this case, the only additional analog signal conditioning hardware needed is a variable attenuator and a warm amplifier.
For the higher frequency 800--1200~MHz readout band, quadrature mixers are added to up and down-convert the baseband signals~\citep{johnson_2016}.

The LEKID readout system uses a heterogeneous architecture, where the real-time signal processing is split between the ROACH-2 board and the readout computer attached to it.
This architecture provides flexibility, allowing us to save time series at full bandwidth ($>$10~kHz) for diagnostic purposes, as well as low pass filtered signals appropriate for the bandwidth of signals expected while observing with a  continuously rotating half-wave plate (HWP) ($<$200~Hz) in a deployed instrument.

\subsubsection{Millimeter-wave sources}
The system to optically characterize the LEKIDs consists of three main components: (i) a variable blackbody source, (ii) an electronic millimeter-wave (MMW) source that can produce broadband or coherent radiation, and (iii) a stepped HWP. A diagram of the optical layout is shown in Fig.~\ref{fig:setup}. 

The blackbody load is a beam-filling piece of Eccosorb absorber (13~mm thick), which is anti-reflection (AR) coated with etched Teflon (380~$\mu$m thick). 
It is weakly thermally coupled to the 2.7~K stage.
The blackbody temperature $T_{\mathrm{bb}}$ can be adjusted from 3--9~K using a resistive heater. 

The MMW source is mounted outside the cryostat.
A WR6 directional coupler splits the source signal. 
Radiation from one port is routed into the cryostat via WR6 waveguide and then launched from a conical horn through the HWP. 
Radiation from the second port is used to continuously measure the power emitted by the source $P_\mathrm{s}$ using a calibrated zero-bias detector (ZBD).
The MMW source can be operated in broadband mode (140--165~GHz) or continuous wave mode~\citep{flanigan_2016a}.
In continuous wave mode, the frequency of the single tone can be swept.

The polarization orientation of the source signals can be rotated with the stepped HWP.
The HWP is sapphire (3.2~mm thick) with fused silica AR coatings (0.28~mm thick). 
The HWP is rotated by a cryogenic motor mounted at 2.7~K controlled by an Arduino-based system at room temperature.
After the HWP, Eccosorb, which is mounted on an aluminum thermal bus, acts as an 11~dB attenuator for the millimeter-radiation.
The Eccosorb therefore acts both as the blackbody source and as an attenuator in the path of the radiation from the MMW source.
The capabilities of this system allow us to measure the detector noise spectra, responsivity, absorption spectra, and polarization selectivity.


\section{Results} 

 \begin{figure*}[t!]
   \centering
   \begin{subfigure}[b]{0.48\textwidth}
   \includegraphics[width=\textwidth]{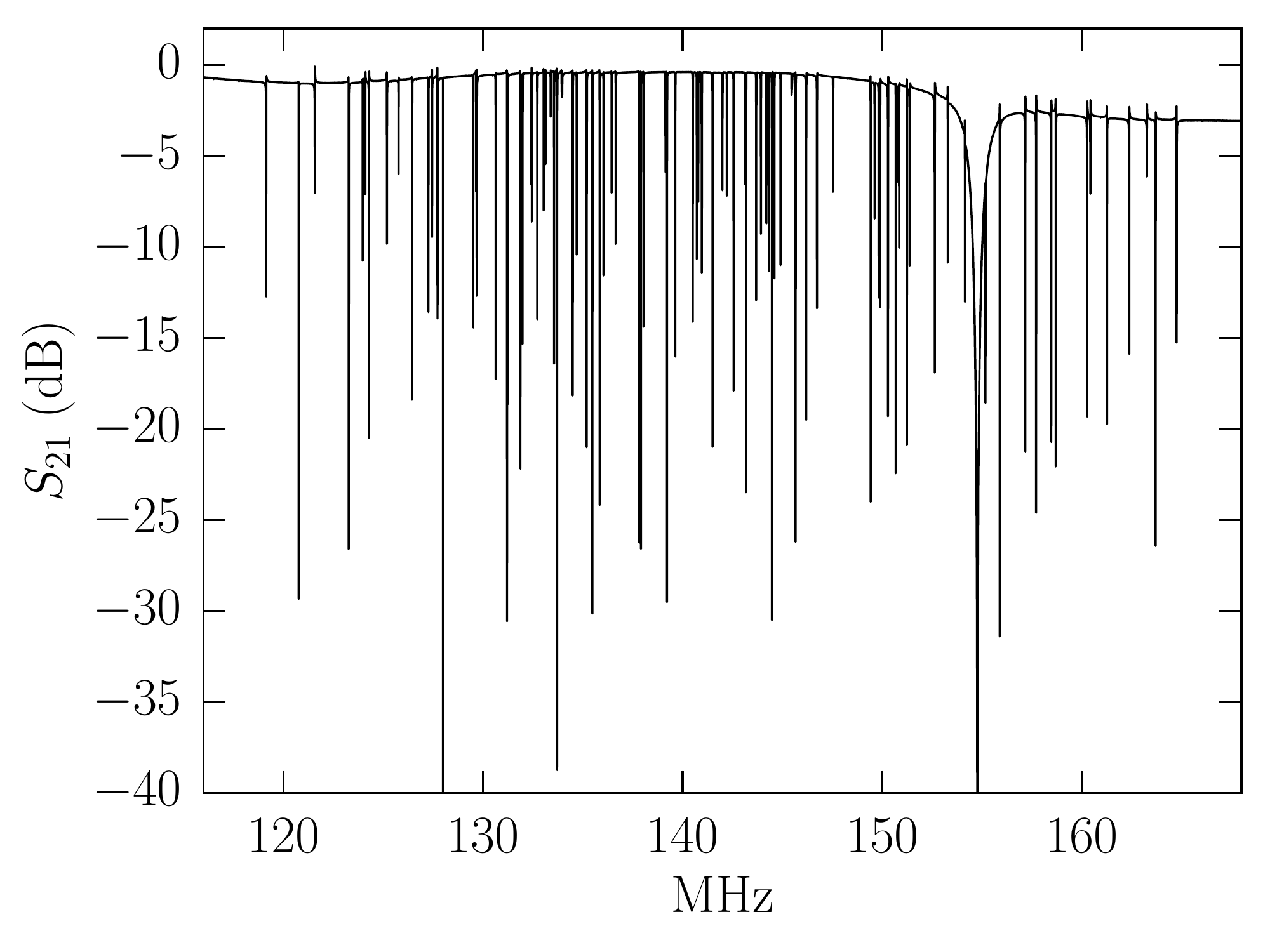}
   \vspace*{-7mm}
   \caption{}
   \end{subfigure}
   \begin{subfigure}[b]{0.48\textwidth}
   \includegraphics[width=\textwidth]{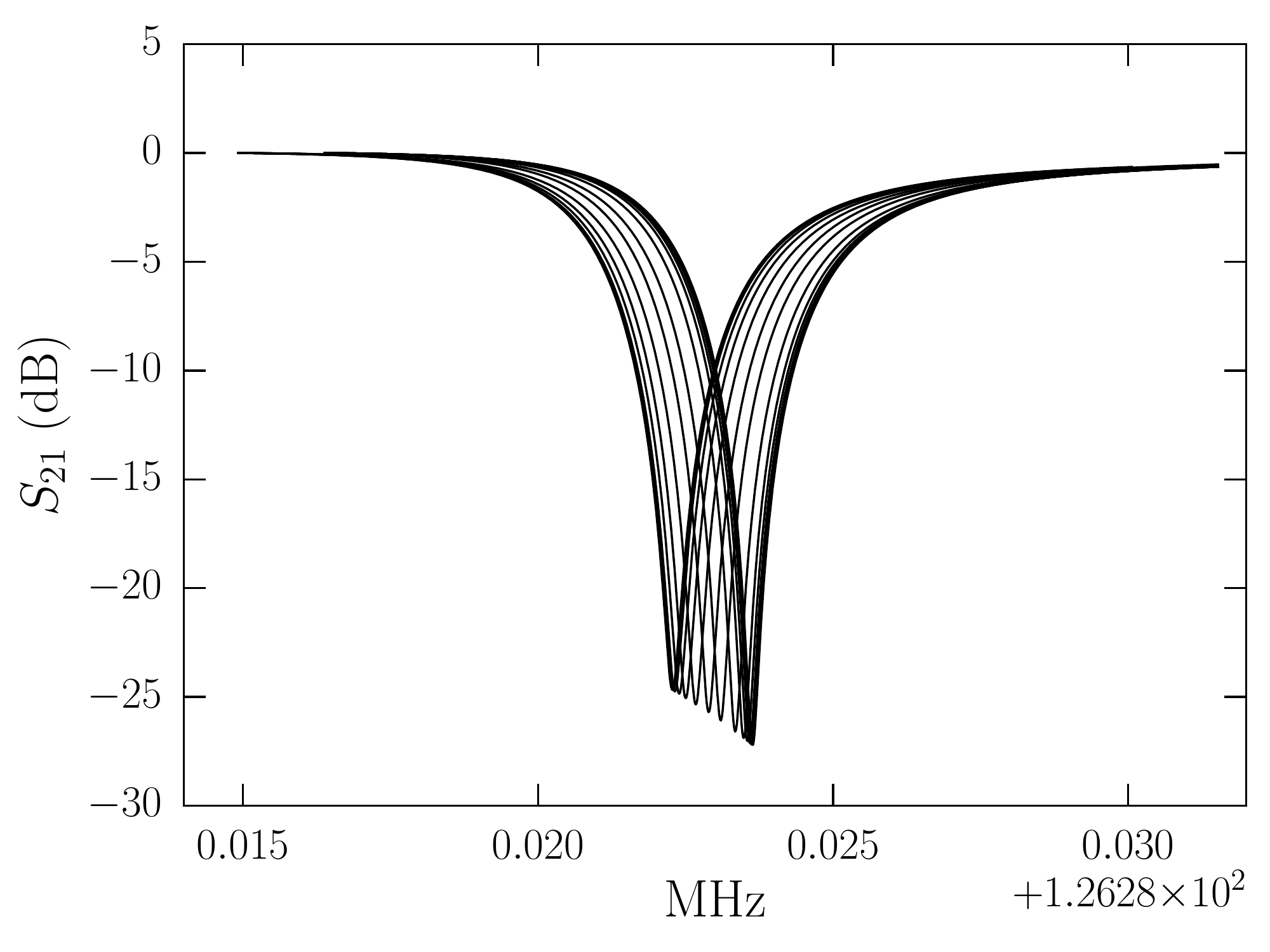}
   \vspace*{-7mm}
   \caption{}
   \end{subfigure}
      \caption{(a) Frequency sweep across the array. Each dip in transmission corresponds to an individual resonator.  The total resonator $Q$ is dominated by $Q_\mathrm{c}$ and is in the range of $10^{4}$ across the array as designed. The internal quality factors $Q_\mathrm{i}$ approach $1 \times 10^{6}$. 
      (b)~Resonance of a single LEKID responding to increasing optical power. The quasiparticle density $n_{\mathrm{qp}}$ increases with optical loading, causing a shift in $f_\mathrm{0}$ and $Q$.}
         \label{fig:s21}
   \end{figure*}

The experiments described in the following section refer to a single 128 resonator array with resonance frequencies that fall between 110--170~MHz. 
We performed a sequence of cryogenic cooldowns in the following configurations: (i) the horn array was removed and an aluminum plate closed the package, (ii) all horn apertures were illuminated by the sources, and (iii) all but one horn aperture was covered. 

\subsection{Resonator characterization}
For the first test, the detectors were enclosed in a sealed box and not illuminated by any of the aforementioned sources. 
The temperature of the array was held at 120~mK. 
We measured $S_{\mathrm{21}}$, the forward scattering parameter, across the array, as shown in Fig.~\ref{fig:s21} and identified the resonance frequencies. 
This $S_{\mathrm{21}}$ measurement revealed 100 out of 128 resonators for a 78\% yield. 

Each resonance is fit to the equation
\begin{align}
\label{eq:s21}
S_{\mathrm{21}} = 1 - \frac{Q}{Q_\mathrm{c}} \left(\frac{1}{1+2\mathrm{j}Qx}\right),
\end{align}
where $Q_\mathrm{c}$ is the complex coupling quality factor, $Q$ is the resonator quality factor, $x$ is the fractional frequency shift (Eq.~\ref{eq:x}), and we have omitted an overall phase term for clarity~\citep{khalil_2012}.
The resonators have $Q_\mathrm{i}$ approaching $10^{6}$ and $Q_\mathrm{c}$ in the range of $10^{4}$ as designed. 

We set the probe-tone power slightly below the bifurcation level to strike a balance between suppressing the amplifier noise and ensuring the detectors remain in a linear regime. 
To determine the bifurcation level of the detectors, we measured $S_{\mathrm{21}}$ and collected time-ordered data (TOD) at different probe-tone attenuations. 
For the dark configuration, the detectors bifurcate at approximately -105~dBm.

In a subsequent cryogenic cooldown, with the horn array installed, we measured the resonator quality factors.
The horn apertures faced a beam-filling 3.4~K blackbody load.
Under these loading conditions the median internal quality factor $Q_\mathrm{i}$ is $3 \times 10^5$ and the median resonator quality factor $Q$ is $3 \times 10^4$.
These quality factors meet the requirements for a 512 multiplexing factor (see Sec.~\ref{sec:detector_design}).


\subsection{Blackbody response}
\label{sec:response}
We measured the responsivity of the detectors $\mathrm{d}x/\mathrm{d}T_{\mathrm{bb}}$ with the blackbody load, where $T_{\mathrm{bb}}$ is the blackbody temperature. 
The absorbed power $P$ can be related to $T_{\mathrm{bb}}$ by integrating the Planck equation over the spectral band.
In the regime above 3~K the relationship between $P$ and $T_{\mathrm{bb}}$ is approximately linear. 
At low $P$ the relationship between $x$ and $P$ is linear. 
At high $P$, $x \propto P^{1/2}$~\citep{flanigan_2016a} and therefore $\mathrm{d}x/\mathrm{d}P \propto P^{-1/2}$. 

\begin{figure*}[h!]
   \centering
   \begin{subfigure}[b]{0.48\textwidth}
   \includegraphics[width=\textwidth]{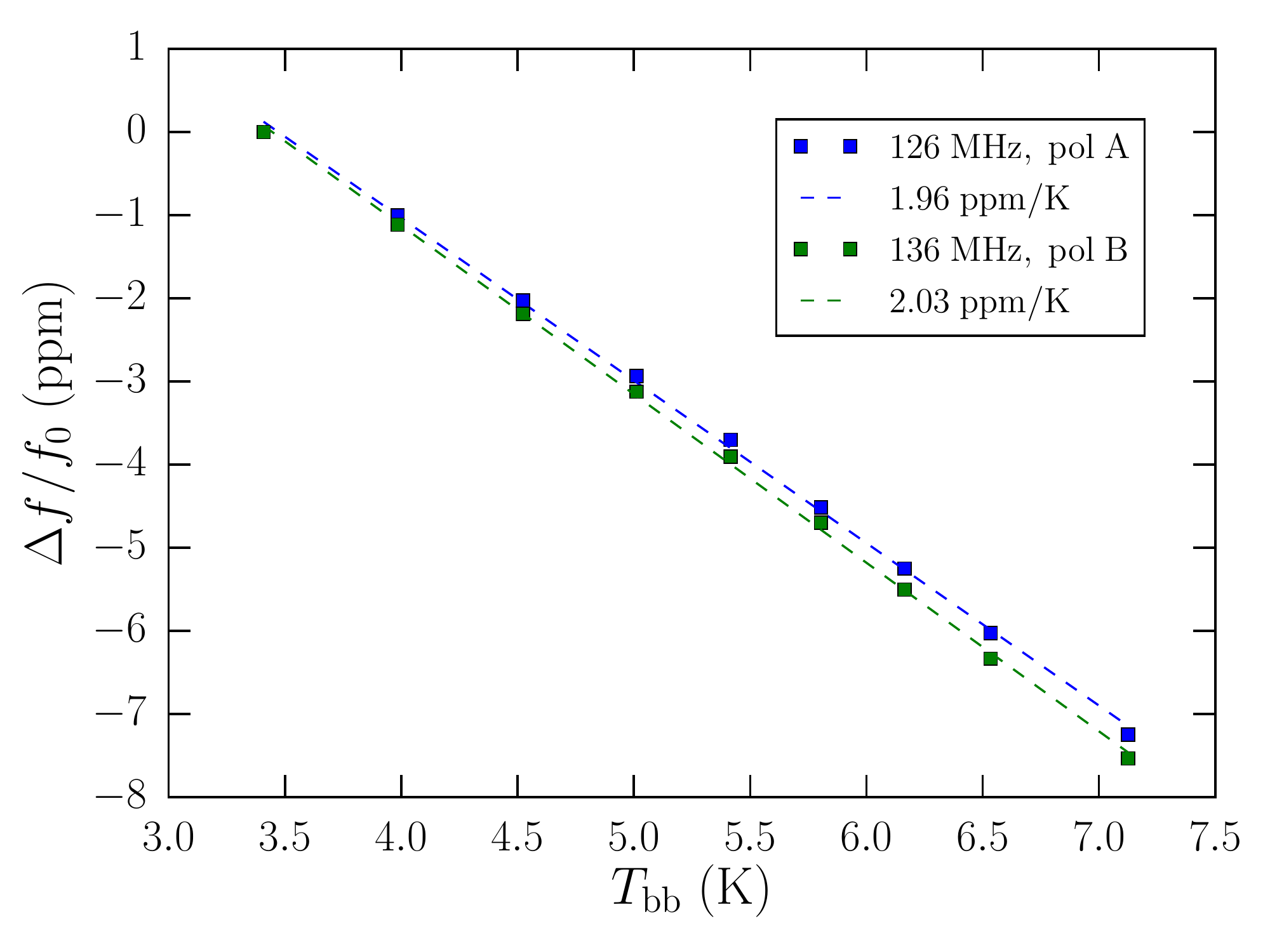}
   \vspace*{-7mm}
   \caption{}
   \end{subfigure}
   \begin{subfigure}[b]{0.48\textwidth}
   \includegraphics[width=\textwidth]{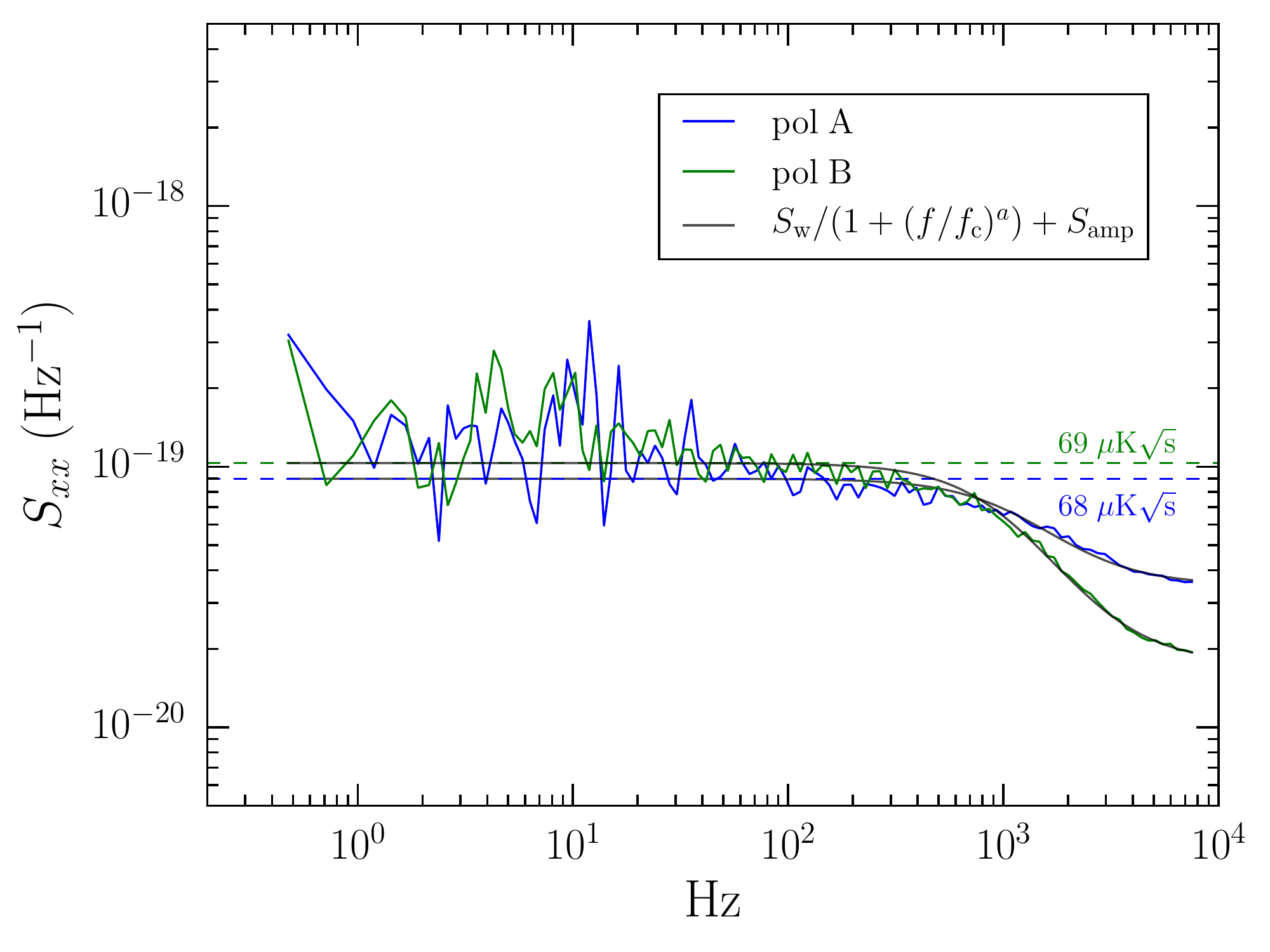}
   \vspace*{-7mm}
   \caption{}
   \end{subfigure}
      \begin{subfigure}[b]{0.48\textwidth}
   \includegraphics[width=\textwidth]{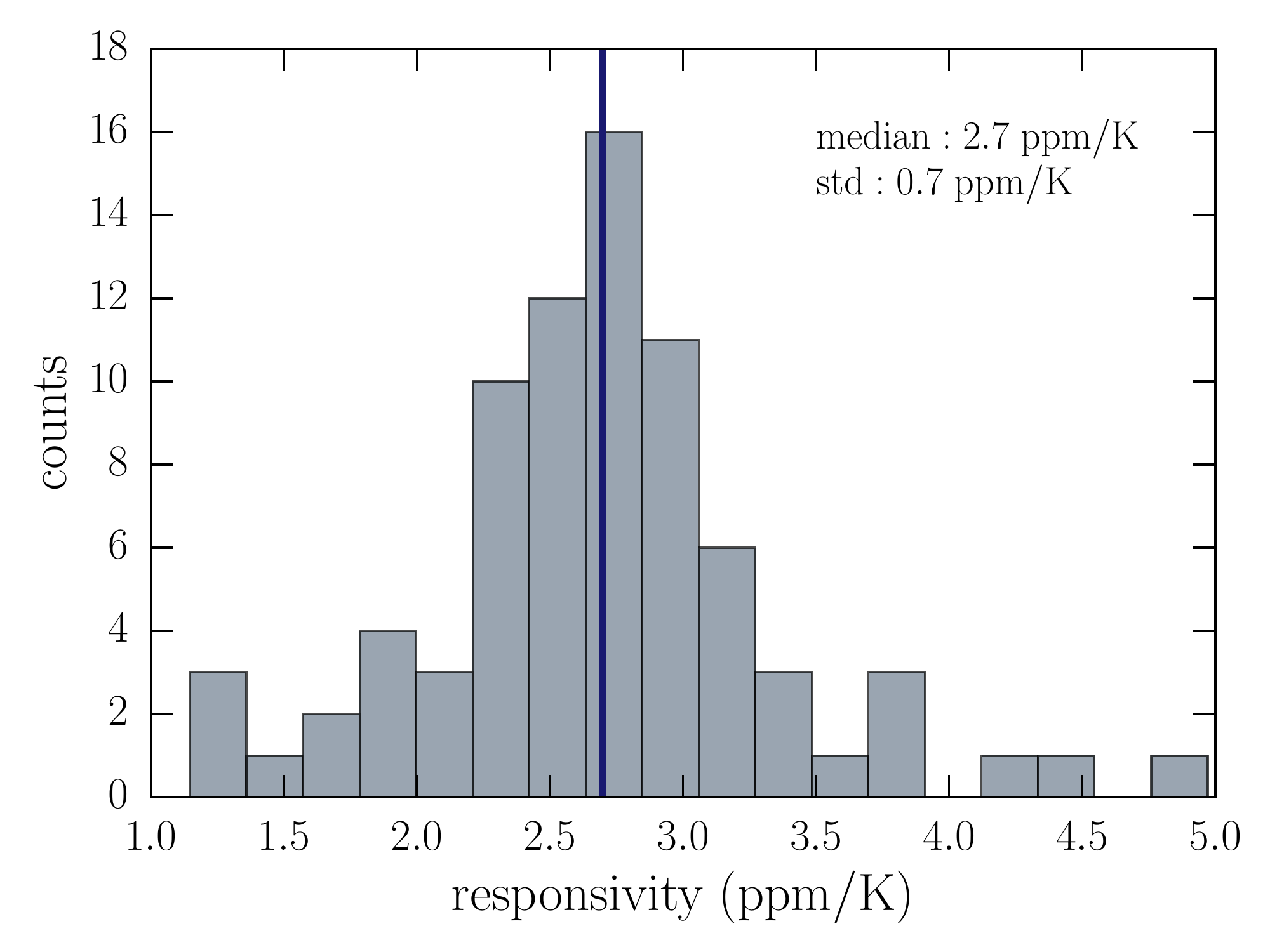}
   \vspace*{-7mm}
   \caption{}
   \end{subfigure}
   \begin{subfigure}[b]{0.48\textwidth}
   \includegraphics[width=\textwidth]{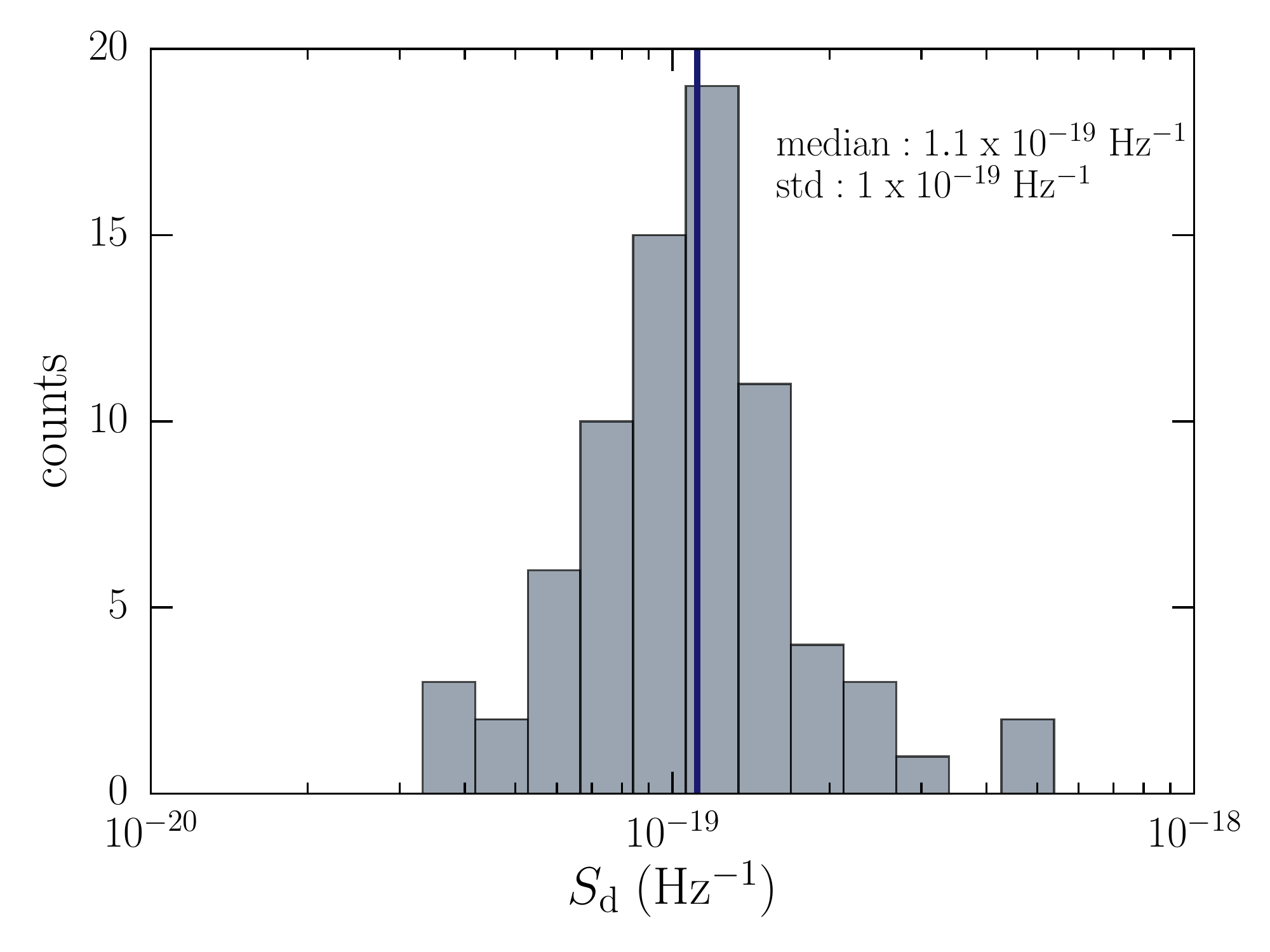}
   \vspace*{-7mm}
   \caption{}
   \end{subfigure}
      \caption{
      (a) Fractional frequency response ($\Delta f/ f_\mathrm{0}$) for a pair of LEKIDs plotted as function of the blackbody load temperature. The response of the two resonators sensitive to orthogonal polarizations is approximately 2~ppm/K and linear.
      (b) Noise spectra for detectors sensitive to orthogonal polarizations under a 3.4~K blackbody load. The fits of the noise spectra are plotted in solid black. The noise in both detectors is similar in shape with the white noise rolling off at approximately 700~Hz. The device noise is plotted as the dashed line and is used in combination with the response of the detectors to calculate the NET. The NETs referenced to 3.4~K for polarization A and B are 68 and 69~$\mu \mathrm{k \sqrt{\mathrm{s}}}$ respectively.
      (c) Histogram of the responsivity of the detectors across the array. The data is for 78 detectors that were read out automatically using our readout software. When read out automatically, the number of detectors with usable data is primarily defined by how many readout tones land sufficiently close to the targeted resonance frequencies without further refinement. 
      (d) Histogram of the device noise level for 78 detectors across the array.}
         \label{fig:pair}
   \end{figure*}

The fractional frequency shift $x$ is plotted as a function of $T_{\mathrm{bb}}$ for both detectors in a single array element in Fig.~\ref{fig:pair}a.
The blackbody temperature is stepped from 3.4--7.0~K.
We find the data for both polarizations is well fit by a line with a slope of approximately 2~ppm/K, which is the responsivity. 
Across the array, the median responsivity is 2.7~ppm/K with a standard deviation of 0.7~ppm/K as shown in Fig.~\ref{fig:pair}c. 
Previous iterations of dual-polarization detectors with straight inductors and smaller inductor volumes had responsivities between 12 and 16~ppm/K~\citep{mccarrick_2016b}.
The increase in inductor volume with the wiggled design has effectively lowered the responsivity and increased the dynamic range (more in Sec.~\ref{sec:range}). 

\subsection{Noise spectra and NET}
\label{sec:noise}
We measured the noise properties of the detectors by recording TOD with the probe-tone frequencies fixed at the resonance frequencies determined by $S_{\mathrm{21}}$ measurements.
The resonator model (Eq.~\ref{eq:s21}) is used to transform the TOD into fluctuations in resonance frequency and dissipation $\delta Q^{-1}$. 
We then calculate the power spectral densities (PSD) for both the fractional frequency and dissipation fluctuations, which we refer to as $S_{xx}$ and $S_{yy}$, respectively. 
In practice we use the fractional frequency data because it is more responsive than the dissipation data. 
The noise spectrum $S_{xx}$ is fit to the model
\begin{align}
\label{eq:noise_model}
 S_{xx} = S_\mathrm{w} \left(\frac{1 + \left(f_\mathrm{k}/f\right)^{\alpha}}{1 + (f/f_\mathrm{r})^{2}}\right) + S_{\mathrm{amp}},
\end{align}
where $S_\mathrm{w}$ is the white-noise level, $S_{\mathrm{amp}}$ is the amplifier noise level, $f_\mathrm{k}$ is the low-frequency noise knee, and $f_\mathrm{r}$ corresponds to the high frequency roll off. 

Typical measured detector noise spectra for both polarizations under a 3.4~K blackbody load are shown in Fig.~\ref{fig:pair}b. 
Thirty seconds of TOD were collected with the sub-kelvin stage temperature regulated but the PTC off to eliminate mechanical vibrations.
The module temperature was stable during the measurement but the radiation environment in the cryostat detectably changed. 
The noise spectra are therefore calculated from TOD detrended with a 2nd degree polynomial.
The white noise levels $S_\mathrm{w}$ for this detector pair are 5 and 9$~\times~10^{-20}~\mathrm{Hz^{-1}}$.
We find that the noise spectra do not show detectable $f^{-\alpha}$ dependence in our measurements. 
TLS noise is expected to have a $f^{-0.5}$ dependence~\citep{noroozian_2009}, so any TLS noise in our devices would have an $f_\mathrm{k}$ below 0.5~Hz.  
At high frequencies, the spectrum rolls off at approximately 700~Hz, which is consistent with the resonator bandwidth $b_\mathrm{r} = f_\mathrm{0} (2Q)^{-1}$. 

The noise equivalent temperature (NET) is calculated as 
\begin{align}
\label{eq:net}
\mathrm{NET} = \sqrt{\frac{S_\mathrm{d}}{2}} \left(\frac{\mathrm{d}x}{\mathrm{d}T_{\mathrm{bb}}}\right)^{-1},
\end{align}
where the device noise level is defined as $S_\mathrm{d} = S_\mathrm{w} + S_{\mathrm{amp}}$. 
The latter two terms are found by fitting Eq.~\ref{eq:noise_model} to the measured noise spectra.
The factor of 1/$\sqrt{2}$ comes from the Nyquist sampling frequency. 
Taking the steps described in Sec.~\ref{sec:response}, we measure $\mathrm{d}x/\mathrm{d}T_{\mathrm{bb}}$.
We find that the NETs are 68 and 69~$\mu \mathrm{K \sqrt{\mathrm{s}}}$ referenced to 3.4~K for the two polarizations as shown in Fig.~\ref{fig:pair}b. 

The median white noise level $S_\mathrm{w}$ across the array is $6.5 \times 10^{-20}~\mathrm{Hz^{-1}}$ and the median device noise level $S_\mathrm{d}$ is $1.1 \times 10^{-19}~\mathrm{Hz^{-1}}$ as shown in Fig.~\ref{fig:pair}c.
When fitting the noise spectra across the array, we use the frequencies above 100~Hz due to spikes at low frequencies in some resonators, which are caused by mechanical vibrations.
To better understand how mechanical vibrations couple to the detector array, we took $x$ TOD from all of the detectors and conducted a spectral coherence analysis.
We calculate the coherence estimator as ${C}_{uv} = \lvert{S_{uv}}\rvert^2 (S_{uu}\ S_{vv})^{-1}$,
where $u$ and $v$ represent different detectors, and $S_{uv}$ represents the cross spectrum of the fractional frequency fluctuations from two detectors.  
We initially found strong coherence in features across the bandwidth of the detectors.
We hypothesized that the thickness of silicon wafer (160~$\mu$m) in combination with the large surface area ($65 \times 65$~mm) caused the array to be sensitive to mechanical vibrations.
To test this hypothesis, we installed the aforementioned spring-loaded pins in the module in an effort to suppress any vibrations in the substrate. 
After the spring loaded pins were installed, we found a large decrease in the coherence. 
However, there is still some detectable coherence below 100~Hz, which means the vibration suppression technique works but still needs further development. 

\begin{figure*}[t!]
   \centering
    \begin{subfigure}[b]{0.48\textwidth}
    	\includegraphics[width=\textwidth]{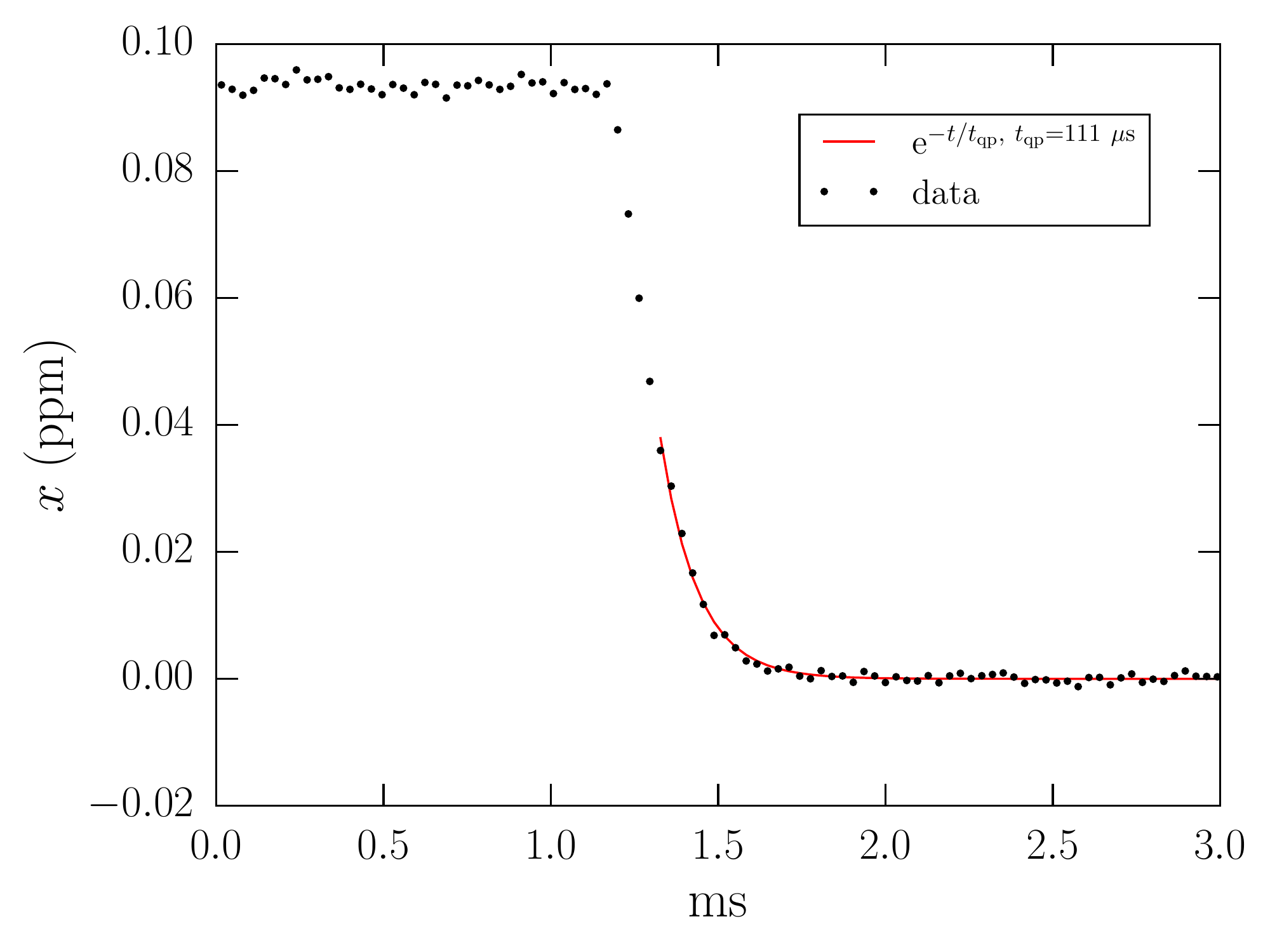}
        \vspace*{-7mm}
        \label{fig:6a}
        \caption{}
    \end{subfigure}
        \begin{subfigure}[b]{0.48\textwidth}
        \includegraphics[width=\textwidth]{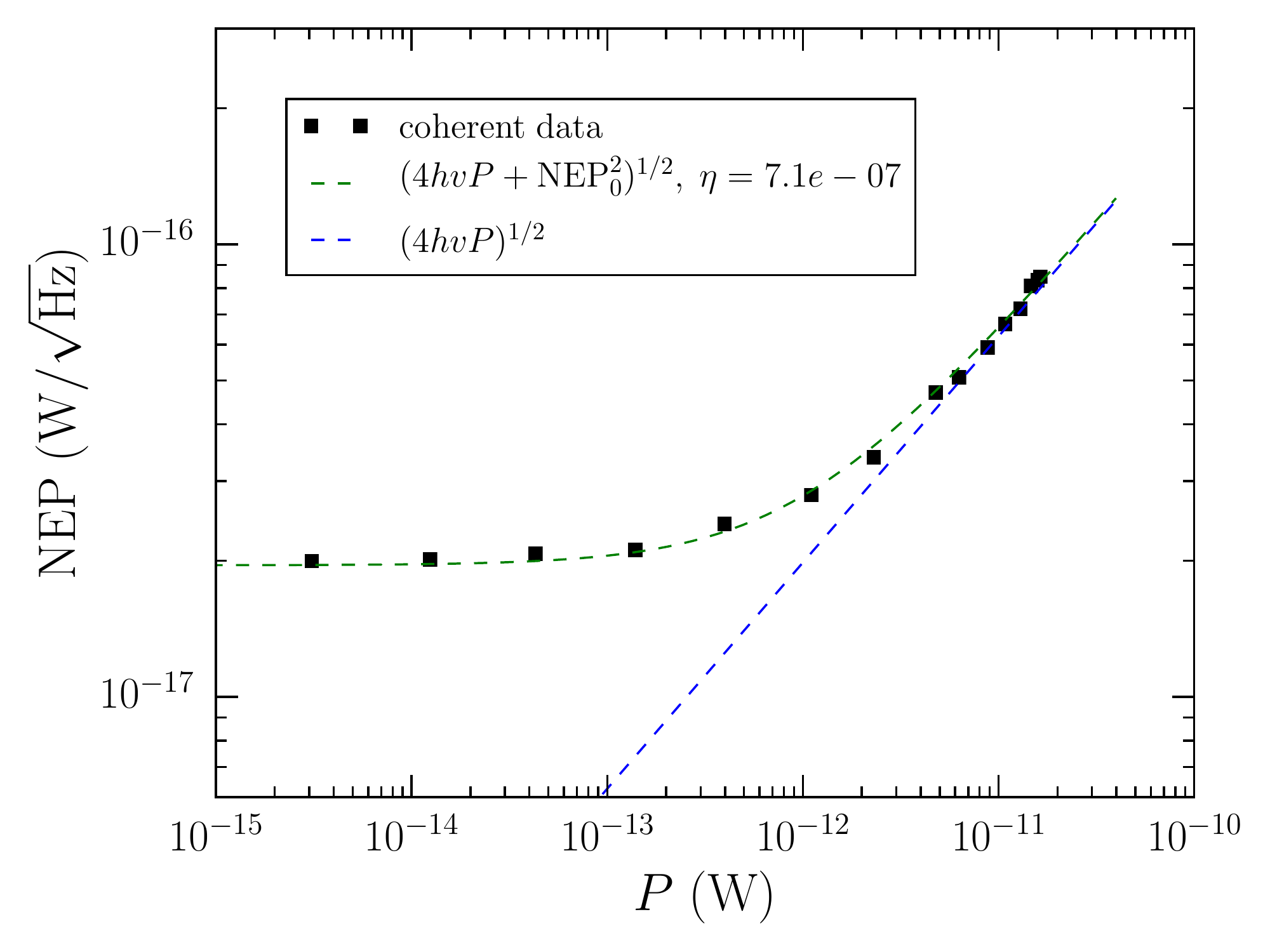}
        \vspace*{-7mm}
        \label{fig:6b}
        \caption{}
    \end{subfigure}
    \caption{(a) Resonator response to a 122~Hz chopped signal. The TOD has been folded and averaged. We use a higher mode resonance ($\sim$~1~GHz) of the LEKID to measure the quasiparticle lifetime. We fit the tail of the falling edge to an exponential model to find the quasiparticle lifetime $\tau_{\mathrm{qp}} = 111~\mu s$. The fit is the red line. 
    %
    (b) Detector NEP as a function of absorbed power.  Plotted in the blue dashed line is the expected photon noise for coherent radiation which is shot noise and GR noise in equal parts. The limiting $\mathrm{NEP}_0$, at the lowest powers, is approximately  $2 \times 10^{-17}~\mathrm{W/ \sqrt{\mathrm{Hz}}}$. The devices are photon noise dominated above approximately 1~pW of absorbed power.
    }
    \label{fig:nep_tau}
    \end{figure*}
\subsection{Quasiparticle lifetime}\label{sec:tau}
The quasiparticle lifetime $\tau_{\mathrm{qp}}$ was measured by taking TOD at a higher-order resonance ($\sim 1$~GHz) while the detectors were illuminated by a constant 3.4~K blackbody background load and the external MMW source. 
It is necessary to use a higher-order resonance because the resonator ring-down time is $\tau_{\mathrm{r}} = (2 \pi\ b_\mathrm{r})^{-1} = Q (\pi f_\mathrm{0})^{-1}$, which is longer than $\tau_{\mathrm{qp}}$ for the 100~MHz resonances.
For this test, the MMW source emits 148~GHz radiation, which transmits through the Eccosorb, and is switched on and off at 122~Hz with a PIN diode. 
The switching time of the PIN diode is fast (10~ns) compared to the LEKID response time, so any time constant associated with the MMW source is negligible. 
The TOD is then folded and averaged, the results of which are plotted in Fig.~\ref{fig:nep_tau}.
The tail end of the falling edge of the averaged TOD is fit to an exponential model in order to estimate $\tau_{\mathrm{qp}}$~\citep{baselmans2016,Baselmans2008}.
The fit yields $\tau_{\mathrm{qp}} = 111~\mu$s.

\subsection{Crosstalk}
We define crosstalk as the response of a detector to a mechanism other than the absorption of radiation from its associated horn.  
The crosstalk is measured with all horns but one covered with aluminum tape, leaving a single array element (one LEKID pair) illuminated. 
The array is illuminated by the same chopped MMW source and background load as described in Sec.~\ref{sec:tau}.
In practice, we quantify the crosstalk as the relative response of each dark detector to that of the illuminated pair.
All the detectors, both illuminated and dark, are simultaneously read out.
At the highest incident power levels the MMW source can emit (where $P \sim 10$~pW) the optical crosstalk is below $-20$~dB, which suggests the RF choke works well.

\subsection{Dynamic range}
\label{sec:range}
We determined the dynamic range of the detectors. 
The detectors were illuminated with both the blackbody source, from which we obtain $x$ as a function of $T_{\mathrm{bb}}$, and the MMW source, from which we obtain $x$ as a function of $P_\mathrm{s}$, where $P_\mathrm{s}$ is the power emitted by the MMW source. 
Thus, $P_{\mathrm{s}}$ can be related to $T_{\mathrm{bb}}$, and we can find an equivalent brightness temperature $T_{\mathrm{s}}$ for the MMW source. 
This is advantageous because the MMW source can emit higher brightness temperatures than the blackbody, which starts to appreciably heat the sub-kelvin stage for $T_{\mathrm{bb}} > 9$~K. 

The quality factor and noise of the detectors are measured as a function of brightness temperature from the MMW source.
Two metrics are used for the dynamic range: (i) the $Q$ remains sufficiently high for the multiplexing requirements and (ii) the detector noise remains clearly elevated above the amplifier noise in the noise spectra. 
The detector noise remains separated from the amplifier noise, up to the highest brightness temperature of the MMW source, which is equivalent to approximately 90~K. 
Similarly, we find that $Q > 1 \times 10^{4}$, the minimum required for the desired multiplexing factor, for all accessible $T_{\mathrm{s}}$. 
For instance, for one resonator, $Q$ changes from $7.0 \times 10^{4}$ to $4.5 \times 10^{4}$, while $Q_\mathrm{i}$ decreases from $9 \times 10^{5}$ to $1 \times 10^{5}$ over the entire $T_{\mathrm{s}}$ range.
This puts a lower limit on the maximum brightness temperature the detectors can observe at 90~K. 


 \begin{figure*}[t!]
   \centering
    \begin{subfigure}[b]{0.48\textwidth}
        \includegraphics[width=\textwidth]{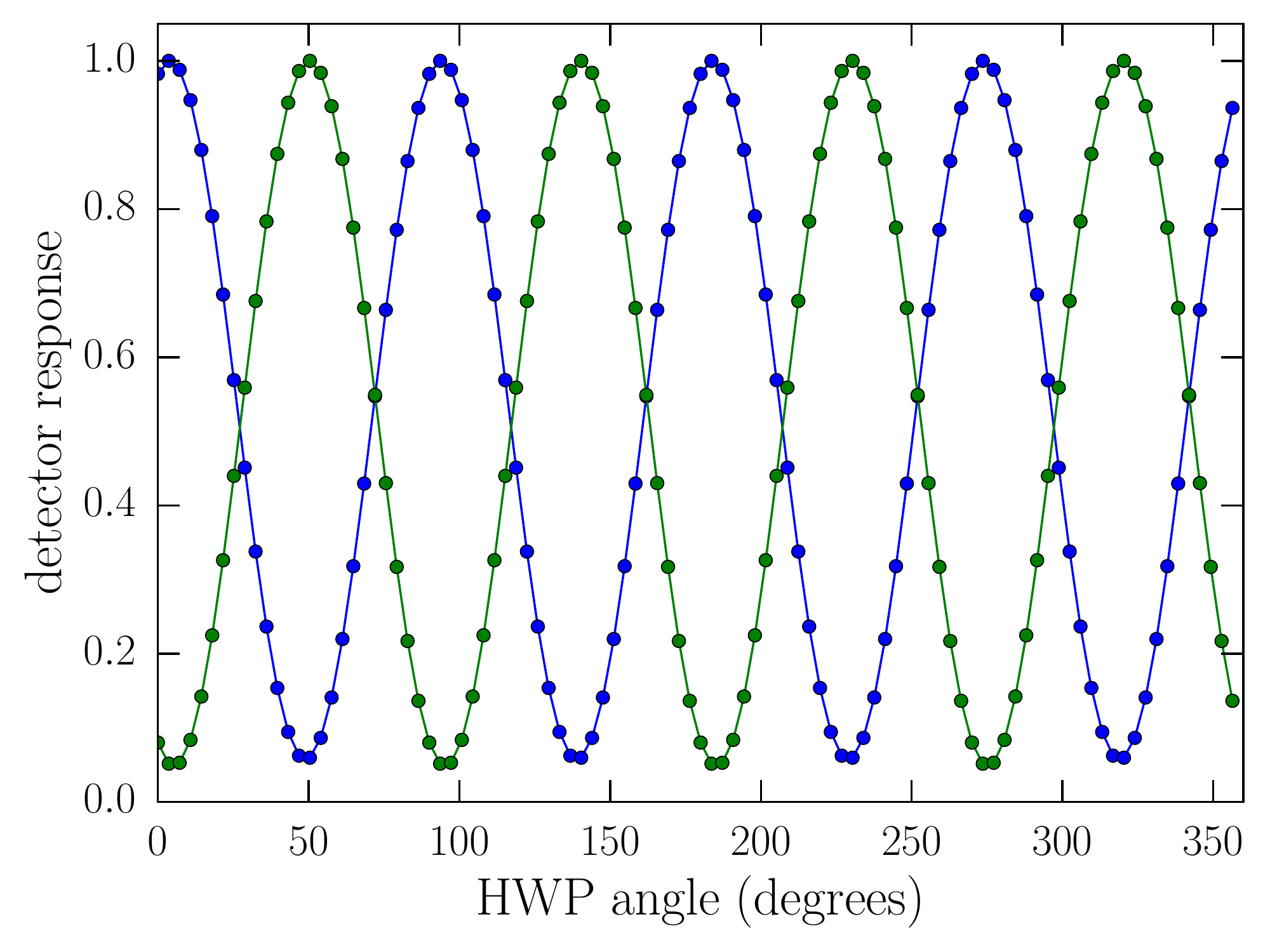}
        \vspace*{-7mm}
        \label{fig:7a}
        \caption{}
    \end{subfigure}
        \begin{subfigure}[b]{0.48\textwidth}
        \includegraphics[width=\textwidth]{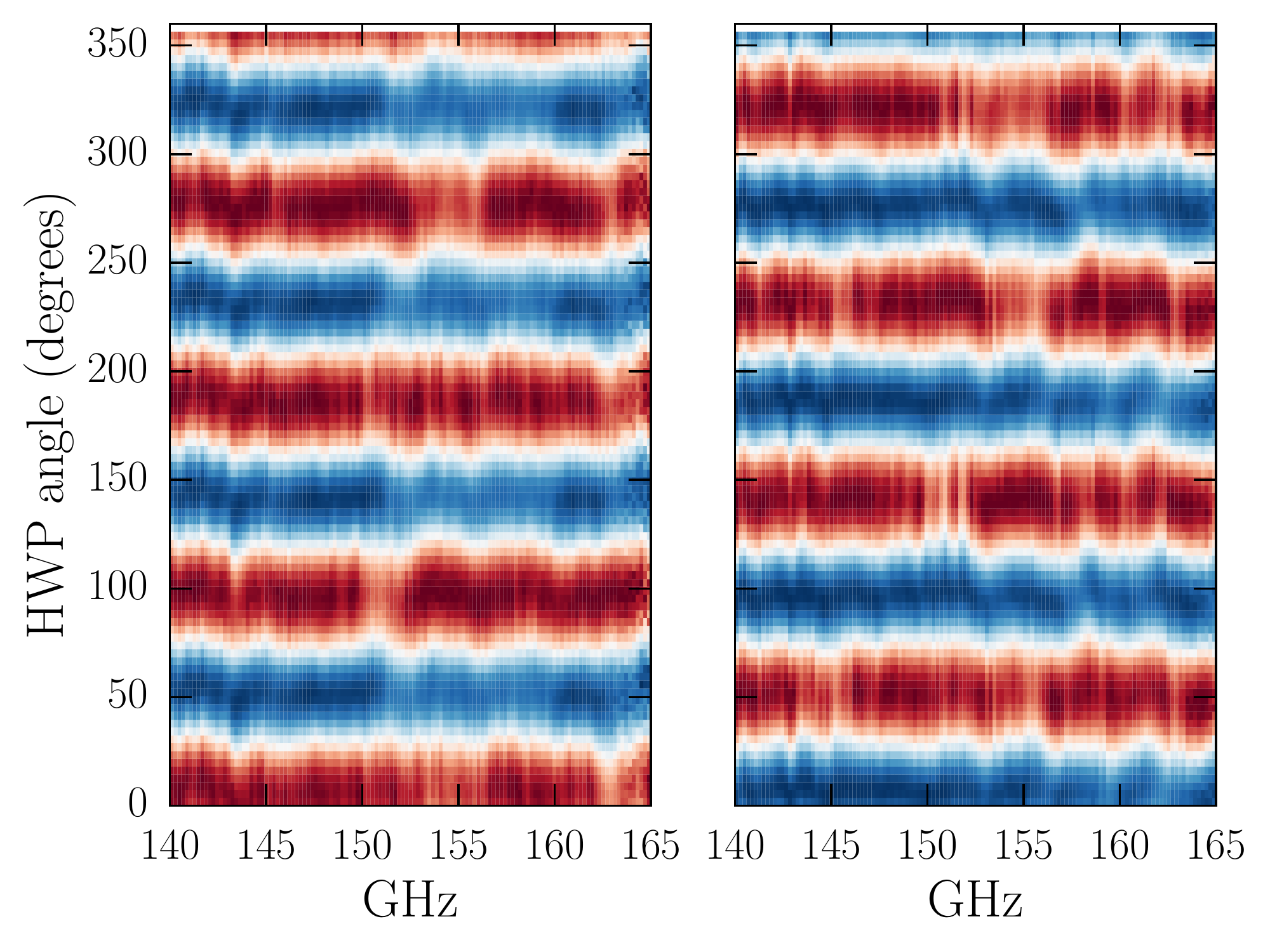}
        \vspace*{-7mm}
        \label{fig:7b}
        \caption{}
    \end{subfigure}
    \begin{subfigure}[b]{0.48\textwidth}
        \includegraphics[width=\textwidth]{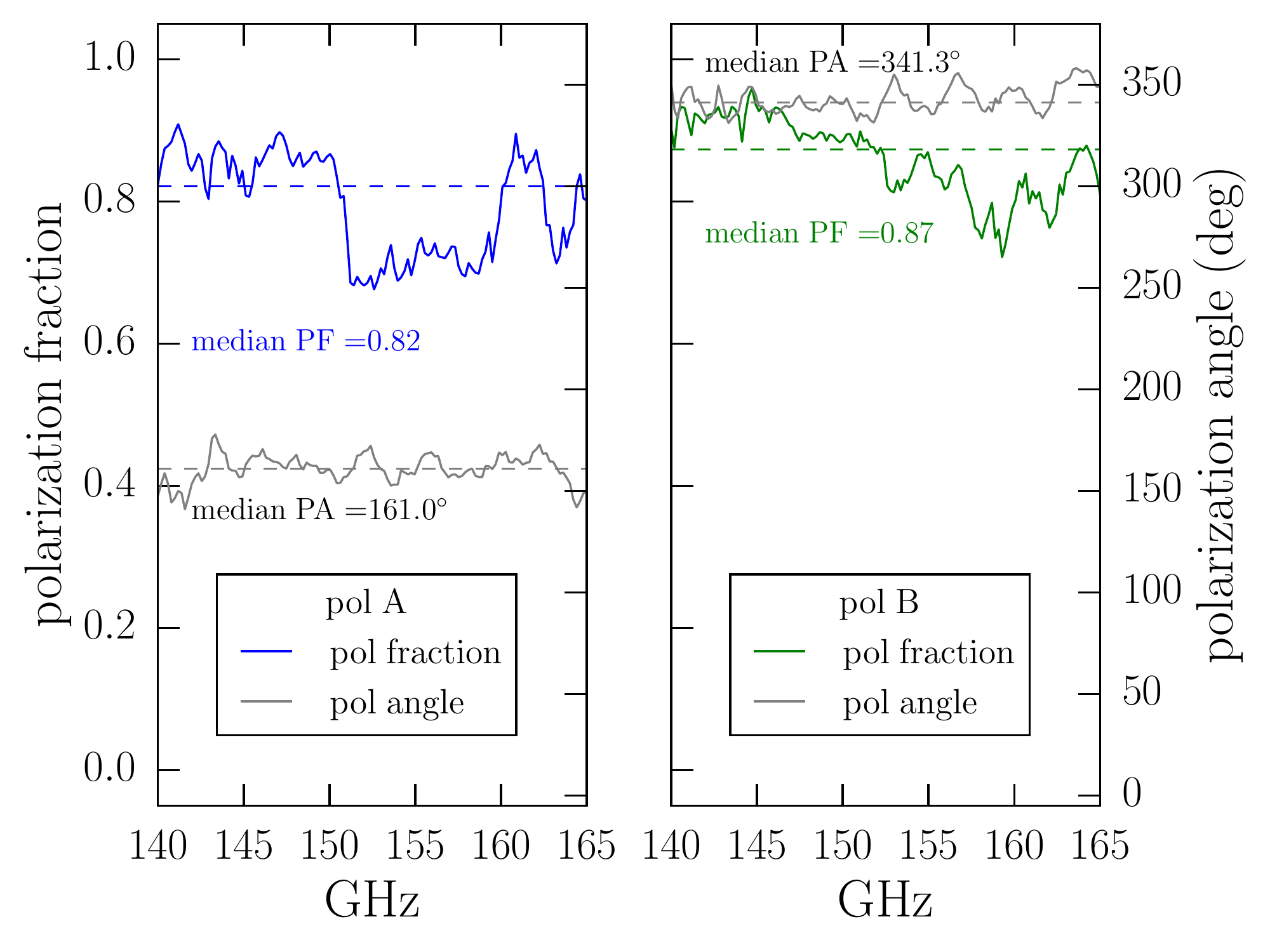}
        \vspace*{-7mm}
        \label{fig:7c}
        \caption{}
    \end{subfigure}
        \begin{subfigure}[b]{0.48\textwidth}
        \includegraphics[width=\textwidth]{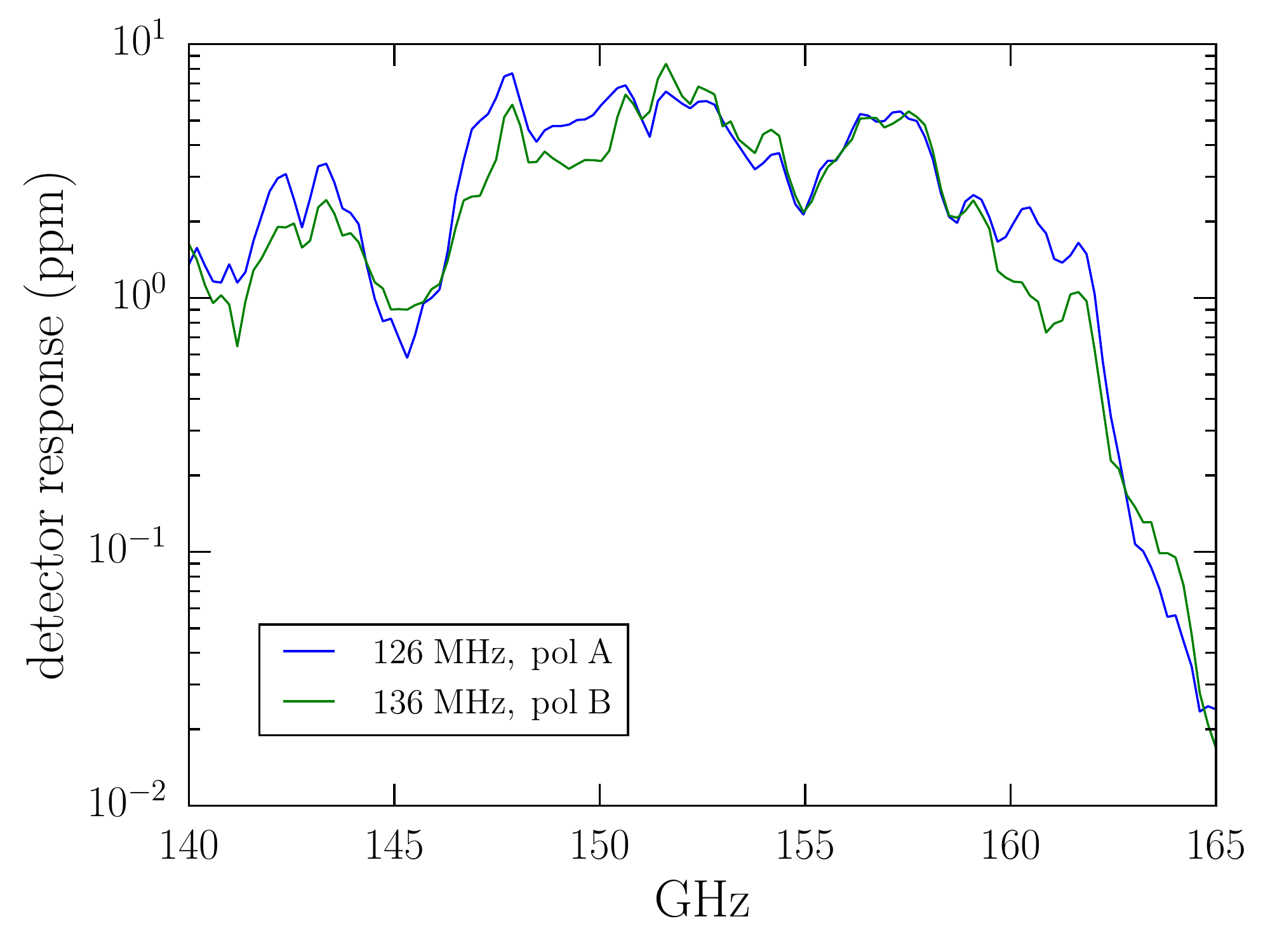}
        \vspace*{-7mm}
        \label{fig:7d}
        \caption{}
    \end{subfigure}
   \caption{
   (a) Normalized detector response as a function of HWP angle for a detector pair with detectors that are sensitive to orthogonal polarizations. The blue points and line correspond to polarization A and the green to polarization B. The data is for a single incident frequency at 147~GHz. The plotted data points have been filtered to show only the DC and fourth harmonic components. 
(b) Detector response as a function of HWP angle and MMW source frequency for the same two detectors. The responses of the polarization A and B detectors are shown in the left and right panels, respectively. The response is peak normalized for each frequency and the colorbar scale goes from blue to red. Fine-scale frequency structure is assumed to be caused by multipath reflections in the experimental setup.
(c) Polarization fraction and angle as a function of frequency for the two detectors. The left panel shows the polarization fraction (PF) in blue and polarization angle (PA) in gray for the polarization A detector. The right panel shows the PF in green and PA in gray for the polarization B detector. The PF is computed by fitting the fourth harmonic and DC components to the data in (b), as described in the text. The relative phase between the HWP response of the two detectors is $179.8\degr$, corresponding to $89.8\degr$ in polarization angle.
(d) Spectral response of the two detectors at the HWP angle giving maximum response for each detector. The rolloff above 160 GHz is due to a filter in the MMW source. Much of the frequency structure is likely due to the MMW source and reflections in the experimental setup. The frequency structure is very similar between the two detectors.
   }
    \label{fig:pol}
    \end{figure*}
    
%

\subsection{NEP}

We measured the absorbed noise equivalent power (NEP) as a function of absorbed power using the method from~\citet{flanigan_2016a}.
The detectors were illuminated through the Eccosorb by coherent radiation at 148~GHz from the MMW source.
The source power $P_\mathrm{s}$ was varied and at each power level we measured the detector noise and the fractional frequency shift $x$. 
We numerically calculated $\mathrm{d}x/\mathrm{d}P_\mathrm{s}$ to determine the desired responsivity. 
Experimentally, the NEP as referenced to the source is calculated as 
\begin{align}
\label{eq:nep}
\mathrm{NEP}_\mathrm{s} = \sqrt{S_\mathrm{w}} \left(\frac{\mathrm{d}x}{\mathrm{d}P_\mathrm{s}}\right)^{-1},
\end{align}
where $S_\mathrm{w}$ is the white noise level obtained by fitting Eq.~\ref{eq:noise_model} to the measured noise spectrum. 

The theoretical NEP in the photon-noise limited case is 
\begin{align}\label{eq:nep_th}
\mathrm{NEP}_\mathrm{t}^2 &= 2 h \nu P +  2 h \nu P (\eta n_\mathrm{o}) + 4 \Delta P/ \eta_{\mathrm{pb}},
\end{align}
where $P$ is the absorbed power, $\Delta$ is the superconducting energy gap, $n_\mathrm{o}$ is the photon occupancy number and $\eta_{\mathrm{pb}}$ is the pair-breaking efficiency of the absorbed radiation. 
The first two terms are inherently due to photons and are respectively called the shot noise and the wave noise. 
Coherent radiation will produce only shot noise. 
The third term is due to the quasiparticle recombination rate and is referred to as recombination noise. 
We rewrite the theoretical NEP as referenced to the source as
\begin{align}
\mathrm{NEP}_\mathrm{s}^2 &= 4 h \nu P_\mathrm{s} / \eta_\mathrm{s} +  2 P_\mathrm{s}^2 /B + \mathrm{NEP}_0^2,
\end{align}
where $\eta_\mathrm{s}$ is the absorption efficiency of the detector as referenced to the MMW source power.   
Here, we have made the substitutions $\eta_{\mathrm{pb}} = 2 \Delta/ h \nu$~\citep{devisser_2015} and $P = h \nu \eta n_\mathrm{o} B$, where $B$ is the bandwidth of incident radiation. 
We have also added in a constant term $\mathrm{NEP}_0$ which refers to the limiting NEP. 

For one representative detector, we fit the coherent data as referenced to the source power $P_\mathrm{s}$ to the expected $\mathrm{NEP}_\mathrm{s}$ model. 
Using $\nu = 148$~GHz, we find $\eta_\mathrm{s} = 5.8 \times 10^{-7}$. 
We use the $\eta_\mathrm{s}$ parameter to calculate the absorbed NEP as a function of absorbed power $P$, which is plotted in Fig.~\ref{fig:nep_tau}.
We see $\mathrm{NEP} \propto P^{1/2}$ when $P > 1$~pW, signifying that the detector is photon-noise limited and detecting shot noise from the coherent radiation. 
The limiting $\mathrm{NEP}$ level is approximately $2 \times 10^{-17}~\mathrm{W/ \sqrt{\mathrm{Hz}}}$, similar to the NEP expected from a 3~K load.

\subsection{Polarization response}
The polarization selectivity of the devices was characterized using a stepped HWP measurement, as is schematically depicted in Fig.~\ref{fig:setup}.
Purely linearly polarized MMW radiation is brought into the cryostat on a WR6 waveguide. 
This signal is then emitted from a conical horn and the emitted beam passes through the HWP, which is mounted directly in front of the horn aperture. 
The HWP is rotated in $3.6\degr$ steps through one full rotation.  
At each orientation we measured the fractional frequency response $x$ and found that the response showed four cycles of a sinusoid for every rotation of the HWP as expected (see Fig.~\ref{fig:pol}a)~\citep{johnson_2007}.
The measured polarization fraction (PF) is defined as
\begin{align}
\label{eq:pf}
\mathrm{PF} = \frac{(x_{\mathrm{max}} - x_{\mathrm{min}})}{( x_{\mathrm{max}} + x_{\mathrm{min}} )} = \frac{I(4f)}{I(0)} = \frac{2 \tilde{I}(4f)}{\tilde{I}(0)}. 
\end{align}
Here, $I(nf)$ corresponds to the intensity of the $n^{\mathrm{th}}$ harmonic and $\tilde{I}(nf)$ corresponds to the intensity of the $n^{\mathrm{th}}$ harmonic as computed by the discrete Fourier transform (DFT).
We are able to calculate the intensity of the harmonics using a DFT as the measurements are taken at evenly spaced intervals over a complete period. 
For a single frequency at 147~GHz, we find the PF is approximately 89\%. 
We expect this PF to be a lower bound on the polarization selectivity as we do not correct for cross-polarization induced in the setup, such as from the HWP or conical horns.

To measure the PF as a function of frequency, at each HWP orientation the incident MMW radiation frequency is swept from 140--165~GHz in 195~MHz steps. 
The results are plotted in Fig.~\ref{fig:pol}b.
We clearly see the two detectors are out of phase by approximately $180^{\circ}$ across the spectral band, which is expected if the detectors are sensitive to orthogonal polarizations. 

The response taken at the HWP angle that produced the maximum response is plotted at a function of incident MMW frequency in Fig.~\ref{fig:pol}c.
The median polarization fraction is 82\% and 87\% for polarizations A and B, respectively.
The detectors are sensitive to orthogonal polarizations and have median responses $179.8\degr$ out of phase corresponding to a polarization angle of $89.9\degr$.

The spectral response of a pair of detectors fed by the same horn is plotted in Fig.~\ref{fig:pol}d.
The maximum response is taken at each incident MMW frequency. 
The response of both polarizations is similar across the entire MMW band. 
It is likely that the small scale structure in the spectral response is a systematic effect caused by the MMW source and reflections in the experimental setup. 


\section{Discussion}
Future CMB experiments will require additional spectral bands and this dual-polarization LEKID design is directly scalable to other frequencies.
While aluminum is sufficient for frequencies of 100--300~GHz, other materials can be used with the same design for both lower and higher spectral bands. 
For instance, materials such as aluminum manganese with a gap energy suitable for photons with frequencies $< 90$~GHz are currently being demonstrated~\citep{jones_2017}.
For higher frequency bands, like FIR, materials such as TiN are routinely used~\citep{hailey_2016,diener_2012,Dober_2016}.
Arrays of our design, scaled and fabricated with different materials, could be used to cover the entire bandwidth necessary to fully characterize the CMB and galactic foregrounds.
The same design with small modifications could also be used for millimeter-wave observations from space.
We expect approximately 0.3~pW of loading in space as compared to approximately 10~pW at ground-based observatories~\citep{johnson_2015}.
The lower loading conditions would require only a change to the inductor volume to tune the response and dynamic range.

In these tests we have demonstrated a multiplexing factor of 128, the number of detectors that can fit on the 100~mm diameter substrate.
When populating a focal plane, up to 4 arrays can be tiled and connected so only a single pair of cables and one ROACH-based readout is needed for a multiplexing factor of 512.
For this multiplexing scheme, each array in the focal plane would have detectors with unique resonance frequencies within the readout bandwidth. 

We are working on a number of related projects.
First, the number of detectors per array is being increased.
For the higher detector density, we have designed and fabricated hex-packed arrays of dual-polarization LEKIDs fabricated on 100~mm silicon wafers with a 4.8~mm pitch.
This detector density allows for 542~resonators per module. 
We are in the early stages of testing these arrays.
Second, we are developing dual-polarization LEKIDs, which could be used for terrestrial imaging.
The terrestrial imaging detectors are the same design as presented in this paper but with a larger absorbing volume achieved by increasing the film thickness.
Third, we are considering ways to use these detectors for other ground-based, millimeter-wave astrophysical studies such as observations of star forming regions and the Sunyaev-Zel'dovich effect~\citep{svoboda_2016, science_book_2016}.

Subsequent iterations of the dual-polarization modules will further optimize the optical coupling scheme.
The horns should either be profiled, the baseline design for these detectors going forward, or corrugated. 
For expediency, conical horns were used in the tests outlined in this paper.
The optical coupling will be improved with either of these horn designs. 
Thermal contractions can cause optical misalignment in the array.
HFSS simulations show misalignment can reduce optical efficiency while also increasing cross-polarization. 
Although we cannot probe the efficacy of the current alignment method (edge alignment) future iterations could use a pin and slot method which may better constrain the motion of the array. 

The detector yield and the uniformity of the resonance frequency spacing could be increased. 
Currently, the detector yield is greater than 75\%, however this should be able to reach close to 100\%. 
Non-uniformities in the device metal could cause resonance frequencies to shift or opens in the resonator circuits.
Some of the 160~$\mu$m thick wafers have cracked during cryogenic cycling, so we are developing ways to make the arrays more mechanically robust by, for example, bonding them to thick handle wafers.

\section{Conclusion}
The dual-polarization LEKID array we have demonstrated in this study is the first for millimeter wavelengths and the first optimized for ground-based CMB polarization studies. 
In this paper, we have presented the design and characterization of the dual-polarization LEKIDs. 
The array of LEKIDs optimized to observe a 150~GHz spectral band is comprised of detector pairs sensitive to two orthogonal polarizations.
The detectors have high internal quality factors which reach $1 \times 10^6$. 
The detectors are shown to have low-noise with an NET < 100~$\mathrm{\mu K}\sqrt{\mathrm{s}}$ under a 3.4~K load for the two polarizations.
The detectors have uniform response between polarizations.
The detectors have a large dynamic range and were not saturated up to a brightness temperature of 90~K. 
We demonstrate that the detectors have low crosstalk, below $-20$~dB. 
We demonstrate that within a single element, the LEKIDs sensitive to orthogonal polarizations have high polarization selectivity for millimeter-wave radiation at 82\% and 87\% across the 140--165~GHz band. 
We show that the detectors are photon-noise limited above 1~pW of absorbed power. 
The entire array is read out on a single coaxial line and the 128 multiplexing factor is higher than commonly used by current CMB experiments. 
The detectors and readout are designed to be straightforwardly increased to a multiplexing factor of 512. 
Additionally, the detectors are fabricated from a single layer of aluminum.
The relatively simple fabrication process is a crucial strength of these devices as the next generation of CMB experiments will need a large increase in detector count.
These results show that the dual-polarization LEKIDs are ready for an on-sky demonstration in a ground-based CMB polarimeter. 

\begin{acknowledgements}
We acknowledge the JPL Microdevices Lab, where the devices were fabricated, ASU, where the horn array was machined, and Cardiff University, where the filters were made. 
This work is partially funded by the Office of Naval Research. 
H.~M. is supported by a NASA Earth and Space Sciences Fellowship.
We thank A.~Kerelsky for taking the photomicrographs in Figs.~\ref{fig:kid_design}b and c. 
We thank the Xilinx University Program for their donation of the FPGA hardware and software tools that were used in the readout system. 

\end{acknowledgements}

\bibliographystyle{bibtex/aa}
\bibliography{bibtex/references}

\end{document}